\documentclass[aps,prb,twocolumn,showpacs,10pt,a4paper,nofootinbib]{revtex4}
\usepackage[utf8]{inputenc}
\usepackage{graphicx,amsmath}
\usepackage{amssymb}
\usepackage{amsthm}
\usepackage{slashed}
\usepackage{braket}
\usepackage{color}

\newcommand{\ee}{\end{align}}

\newcommand{\pref}[1]{(\ref{#1})}
\newcommand{\oncite}[1]{Ref.~\onlinecite{#1}}
\newcommand{\oncitep}[1]{Ref.~\onlinecite{#1}. }

\newcounter{mycount}

\newcommand\ie {{\it i.e. }}

\newcommand\etcp{{\it etc.. }}
\newcommand\etck{{\it etc., }}

\newcommand\half{\frac 1 2 }

\newcommand{\av}[1]{\langle #1\rangle}

\newcommand\psid {\psi^\dagger}
\newcommand\psib {\bar\psi}
\newcommand\gfem{\gamma_5}
\newcommand\vth{\vartheta}
\newcommand\vph{\varphi}
\newcommand\bxi{\xi^\dagger}

\newcommand{\pij}{$\pi$-junction }

\newcommand {\be}[1]{
      \begin{align} \mbox{$\label{#1}$}  }
\newcommand {\ben}[1]{
      \begin{eqnarray} \mbox{$\label{#1}$}  }

\begin{document}
\title{Topological aspects of $\boldsymbol{\pi}$ phase winding junctions in superconducting wires}
\author{Christian Sp\aa nsl{\"a}tt$^1$, Eddy Ardonne$^1$, Jan Carl Budich$^{2,3}$, and T.H. Hansson$^1$}
\affiliation{$^1$Department of Physics, Stockholm University, SE-106 91 Stockholm, Sweden}
\affiliation{$^2$ Institute for Theoretical Physics,
University of Innsbruck, 6020 Innsbruck, Austria}
\affiliation{$^3$ Institute for Quantum Optics and Quantum Information, Austrian Academy of Sciences, 6020 Innsbruck, Austria}

\date{\today}
\begin{abstract}
We theoretically investigate Josephson junctions with a phase shift of $\pi$~in various proximity induced one-dimensional superconductor models. One of the salient experimental signatures of topological superconductors, namely the fractionalized $4 \pi$~periodic Josephson effect, is closely related to the occurrence of a characteristic zero energy bound state in such junctions.
We make a detailed analysis of a more general type of $\pi$-junctions coined ``phase winding junctions'' where the phase of the order parameter rotates by an angle $\pi$ while its absolute value is kept finite. Such junctions have different properties, also from a topological viewpoint, and there are no protected zero energy modes.  We compare the phenomenology of such junctions in topological ($p$-wave) and trivial ($s$-wave) superconducting wires, and briefly discuss possible experimental probes. Furthermore, we propose a topological field theory that gives a minimal description of a wire with  defects corresponding to $\pi$-junctions. This effective theory is a one-dimensional version of similar theories describing Majorana bound states in half-vortices of two-dimensional topological superconductors.

\end{abstract}
\maketitle

\section{Introduction}

In a 2001 paper, Kitaev predicted the existence of unpaired Majorana zero modes (MZM) localized  at the ends of a proximity effect induced one dimensional (1D) $p$-wave superconductor \cite{Kitaev2001}. The Bogoliubov deGennes (BdG) mean field Hamiltonian of this ``Kitaev chain" is distinguished from a trivial gapped 1D system by a $\mathbb Z_2$-invariant. This topological invariant can be expressed in terms of the Pfaffian of the Bloch-Hamiltonian in the Majorana representation. In the more recently established periodic table of topological states \cite{Schnyder2008,KitaevPeriodic,RyuLudwig}, this invariant is located in the column for dimension $d=1$~in the row for symmetry class D , \ie the class of superconductors without any additional symmetries\cite{AltlandZirnbauer}. 

A single channel nanowire with Rashba spin orbit coupling, in proximity to a bulk $s$-wave superconductor, and subject to an external magnetic field, has been one proposal for an experimentally viable realization of the Kitaev chain \cite{SauTSC,OppenTSC}. A different approach taken is a magnetic impurity chain on top of a superconductor \cite{BeenakkerMagnetic,MagneticTheory,vonOppenShiba,MajoranaMagnetic}.

Due to their charge-neutrality and non-magnetic nature, the unpaired MZMs are not easy to detect. The two main proposed signatures are a zero bias anomaly when the wire is coupled to a normal metal lead, and an anomalous $4\pi$-periodic Josephson effect. Experimental evidence for  the zero bias anomaly has been reported by several experimental groups \cite{LeoMaj, LarssonXu, HeiblumMaj}. However, it is fair to say that alternative explanations for robust zero bias resonances, not related to MZMs, have also been proposed\cite{AltlandZero,LeeZero}. So, in spite of a huge experimental effort, there is still no uncontested experimental realization of a 1D topological superconductor. The search for alternative observable signatures of this state thus remains a key challenge. 

In this paper we investigate  the spectroscopy of sub-gap modes in different types of Josephson junctions in some detail, and ask to what extent this might provide such an alternative signature.
Apart from the frequently considered junctions, in which the order parameter changes sign by going through zero, we also consider junctions for which the phase of the order parameter winds, while the amplitude stays constant.
The sub-gap modes in these junctions can, at least in principle, be detected by standard probes sensitive to the density of states, and in particular scanning tunneling spectroscopy. Since ordinary $s$-wave superconductors can also have sub-gap modes in Josephson junctions, we want to identify spectral features that are specific to the Kitaev chain.

We note that several other studies, complementary to ours, have investigated various aspects of Josephson junctions in topological wires \cite{Ojanen,Arturo,FermFrac}.

The $4\pi$-periodicity of the Josephson effect occurring in a junction between two Kitaev chains was pointed out  already in \oncite{Kitaev2001} (see, e.g. \oncite{FuKaneFourPi} for a detailed discussion). Closely related to this $4\pi$-Josephson effect is a characteristic  level crossing between two sub-gap states associated with a change in the fermion parity of the many body ground state.
This level crossing is accompanied with a fermionic zero energy state localized in the junction region. Here, we study the physics of such junctions in both $s$- and $p$-wave paired wires from a topological perspective, focusing in particular on the nature of the previously mentioned (Dirac) zero mode located at a $\pi$-junction. We recall how the level crossing at phase $\pi$ is protected by an additional pseudo time reversal symmetry (PTRS) which is present in Kitaev's minimal model \cite{Kitaev2001} for the Majorana wire if the pairing field is real (up to a constant phase). This additional symmetry, which is well known to refine the $\mathbb Z_2$~parity to an integer winding number \cite{Ryu2002,SauBDI}, also protects the localized zero mode in the junction region. A major part of our present work is devoted to the study of the more general case where the phase of the superconducting order parameter is allowed to wind in the complex plane in the junction region, thus locally breaking the PTRS. We compare the properties of the \pij in the topologically non-trivial $p$-wave case with those in the trivial $s$-wave case. Even in $s$-wave superconductors, there can still be localized sub-gap modes at a Josephson junction, but  there is no protected zero mode. 

Although  Kitaev's original lattice model can be solved numerically for rather large systems and arbitrary junction profiles, it is nevertheless interesting to verify the presence of the sub-gap modes, and in particular the zero mode, in the junction by analytical means. To achieve this, we linearize the spectrum around the Fermi points to obtain a Luttinger model, augmented with anomalous, charge non-conserving terms, which is  essentially equivalent to the Su-Schrieffer-Heeger model for polyacetylene\cite{Su:1979ut}, with the Josephson junction playing the role of the famous domain wall soliton\cite{Takayama:1980zz}. This allows us to find an analytical solution for the zero mode, and also, for a special order parameter profile, for the full sub-gap spectrum. Although the topological properties of this  linearized model differ from those of the original Kitaev chain, we present both theoretical and numerical arguments for them describing the same physics. 
First we compare with  an alternative linearized model (called below the "V-shape model") which {\em does} have the same topology as the  Kitaev chain. Since this model differs from the first linearized model only at high ($\sim \Delta$) energies it gives theoretical support for our claim that the extended Luttinger model indeed describes the low energy features of the Kitaev chain. Secondly, the analytical results from this  model agrees extremely well with the numerical results obtained by directly diagonalizing the Kitaev chain.

Experimentally, the most obvious way to induce a junction in the wire such that the order parameter  changes sign, is by proximity effect from a bulk superconductor with a real, sign changing order parameter already present - this is the original scenario considered by Kitaev. In such a junction, it is natural to assume that the induced order parameter in the wire remains real also in the junction region, and thus has to vanish at some point. An alternative way to introduce a junction is to place the wire on top of a bulk superconductor through which a current is flowing between two external leads placed below the wire. The resulting phase gradient is, by proximity, also present in the wire. The resulting "phase winding junction" violates the PTRS, and the zero energy state is transformed into a finite energy sub-gap state.

When discussing topological phases, it is interesting to ask what is the minimal model that will encapsulate the topological properties of the phase, and in particular those of the elementary excitations. Important examples are the Chern-Simons theories describing various Quantum Hall liquids\cite{wen1995topological}, and the BF theories describing superconductors and topological insulators\cite{hansson2004superconductors,cho2011topological,chan2013effective}.
In the present case, the elementary excitations carrying topological charge are widely separated $\pi$-junctions at fixed positions, and we show that  the  linearized model, in the background of these $\pi$-junctions can be mapped onto a Dirac equation with a Goldstone-Wilczek type mass term\cite{goodwill}. We take this as a starting point for constructing an effective topological field theory describing the solitons and their associated zero modes, and comment on similar attempts in the case of the 2D topological superconductor.

This article is organized as follows: 
In the next section we first define the models that we shall study.
In section \ref{sec:soliton} we study junctions with a real order parameter for the different models and with both analytical and numerical approaches. Section \ref{sec:phasewindingjunctions} contains a similar analysis for the phase winding junctions with constant absolute value of the order parameter, but in this case we have to rely more heavily on numerics. Section \ref{sec:exp} briefly discusses possible experimental configurations to study the physics of topological $\pi$-junctions, and finally, in section \ref{sec:effectivetheory} we construct the topological field theory referred to above.  We end with a few concluding remarks. Some technical points, and in particular a discussion of the 
rather subtle $k$-space topology of the linearized models, are put in appendices.



\section{Models } 
\label{sec:models}
To set the stage for our analysis, we here first define the various models for the
superconducting wires studied below.

\subsection{The $p$-wave wire}
The Hamiltonian for a spinless (or spin polarized) 1D $p$-wave superconductor can be written as
\begin{align}
\label{contsc}
H_p & = \int dx\, {\mathcal H}_p =  \int dx\, [(\psi^{\dag}(-\frac{\partial_x^2}{2m}-\bar{\mu}) \psi \,    \notag  \\
 &+\Delta_p (x)\psi(-i\partial_x)\psi + \Delta_p^*(x)\psi^{\dag}(-i\partial_x)\psi^{\dag} ]  \ ,
\end{align}
where  $\psi$ is a fermionic field (for simplicity we sometimes suppress  the $x$-dependence), $\bar{\mu}$ is the chemical potential and
$\Delta_p(x) = \Delta(x)/k_F$ is the dimensionless $p$-wave superconducting order parameter. The order parameter, $\Delta(x)$ is defined such that, for constant $\Delta$, the  energy gap is $2\Delta$.

By discretizing the Hamiltonian \eqref{contsc} we get the Kitaev chain model\cite{Kitaev2001} 
\begin{align}
\label{kitaevstart}
H_K &= \sum_{j = 0}^{N-1} ( -t(a_j^\dag a_{j+1} + a_{j+1}^\dag a_j ) \notag \\
&+ \Delta_{j}a_j a_{j+1} + \Delta_{j}^* a_{j+1}^\dag a_j^\dag - \mu(a_{j}^\dag a_{j} -\frac{1}{2}) ) \, .  
\end{align}
Here, the $a_i$ are (spin polarized) fermion operators,
and we have set the lattice parameter to unity for simplicity. The hopping parameter is denoted by $t$, $\mu$ is the chemical potential and $\Delta_{j}$ is the superconducting order parameter which can be position dependent. 
These  parameters are related to those in the  continuum model by $t = 1/(2m)$, and  $\mu = \bar{\mu} - 2t$.

To write the $H_K$ in momentum space (assuming constant $\Delta$), we
introduce the Nambu spinor $\Psi^\dagger_k = (a^\dagger_k , a_{-k})$, in   terms of which, 
$$H_K =\sum_k \Psi^\dagger_k \mathcal{H}_K(k) \Psi_k,$$
with $\mathcal{H}_K(k)$ given by
\begin{align} \label{kitaevbdg}
\mathcal{H}_K(k) =&
\bigl(-\mu/2 - t \cos(k) \bigr) \tau_z - {\rm Re}(\Delta) \sin(k) \tau_y \\ &
+ {\rm Im}(\Delta) \sin(k) \tau_x, \nonumber
\end{align}
where the Pauli-matrices $\tau_i$ act in the particle-hole spinor space.
It is known\cite{Kitaev2001}, that for a constant order parameter, \ie $\Delta_{j} = \Delta$, the Kitaev chain resides in a topological phase when $\Delta \neq 0$ and $|\mu|<2|t|$.

\subsection{The $s$-wave wire}
As discussed in the introduction, we will compare the results for
the topological wires with their topologically trivial, $s$-wave paired, counterparts. 
These trivial wires are described by the continuum Hamiltonian,
\begin{align}
\label{sHam}
H_s &=  \sum_{\sigma = \uparrow,\downarrow} \int dx (\psi^{\dag}_\sigma(-\frac{\partial_x^2}{2m}-\bar{\mu}) \psi_\sigma
 \\ \nonumber &
+ \int dx (\Delta_s (x) \psi^\dagger_{\uparrow} \psi^\dagger_{\downarrow} 
+ \Delta^{*}_s (x) \psi_{\downarrow} \psi_{\uparrow} )),
\end{align}
where $\sigma$ is the spin index,
$\psi_\uparrow$, $\psi_\downarrow$ are fermionic fields,
$\bar{\mu}$ is the chemical potential and
$\Delta_s(x)$ is the position dependent $s$-wave order parameter.


\subsection{Two linearized models}
\label{sec:linear}
To capture the behavior of the above models close to the Fermi energy, we expand $\psi$ into fields
containing only low energy degrees of freedom. We consider two different ways of doing this, which
give the same low-energy physics, but differ in their topological properties.

\subsubsection{Luttinger like model}

\begin{figure}[bt]
\includegraphics[width=\columnwidth]{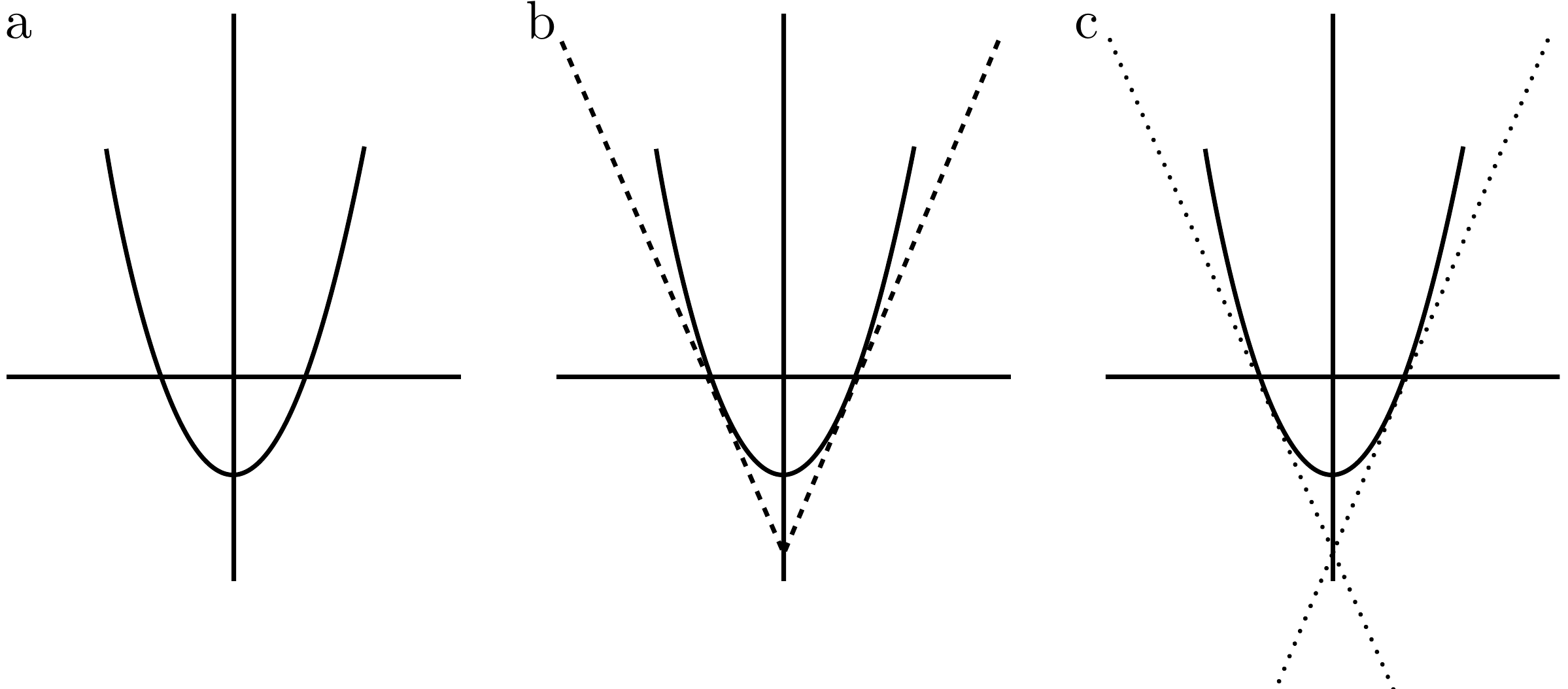}
\caption{Schematic dispersion relations for the free fermion models corresponding putting
$\Delta = 0$ in
(a) $\mathcal{H}_p$ solid line,
(b) $\mathcal{H}_{\rm v}$ dashed line, and
(c) ${\cal H}_{\rm Lin}$ dotted lines.}
\label{fig:linearize}
\end{figure}

There is a standard way to linearize that is illustrated in Fig.~\ref{fig:linearize}(c), where the parabolic band is replaced by a Dirac like dispersion relation.
Just as in the Luttinger model, we have extended the spectrum by adding unphysical "positron" states. In the Luttinger model, 
a gap can be opened by $2 k_F$ processes that scatter electrons between the two Fermi points. In our case a gap is opened by
charge non-conserving processes that creates or destroys a Cooper pair formed by two electrons at different Fermi points.

Formalizing this argument we first define, 
\begin{align*}
\psi = \frac{1}{\sqrt{2}}(e^{i k_F x}\varphi_+ + e^{-i k_F x}\varphi_- )\, ,
\end{align*}
where, $k_F \equiv \sqrt{2m\bar{\mu}}$ is the Fermi momentum, and $\varphi_+$ and $\varphi_-$ are right and left moving fermion fields respectively. Inserting this expression into \eqref{contsc}, neglecting terms $\sim e^{\pm 2i k_F}$,  we obtain 
\begin{align}
\label{linsc}
H_{\rm Lin} &=  \frac{1}{2}\int dx (-iv_F \varphi_+^\dag\partial_x \varphi_+ + iv_F \varphi_-^\dag\partial_x \varphi_- + \notag \\
& +2(\Delta(x)\varphi_- \varphi_+ + \Delta^*(x) \varphi_+^\dag\varphi_-^\dag)) \, , 
\end{align}
where the Fermi velocity is $v_F = k_F/m$.
The quadratic dispersion, $\epsilon (k) = \frac{k^2}{2m} - \bar{\mu}$, is thus effectively replaced by two bands,
corresponding to the right and left moving linearized fermionic fields, with
dispersion relations $\epsilon_\pm (k) = \pm v_F k - \bar{\mu}$. In terms of the momentum
$q$ relative to the respective Fermi momenta, this reads $\epsilon_\pm (q) = \pm v_F q$.
The superconducting order parameter couples these right and left moving fermions.
By introducing the spinor $\Psi^\dagger = ( \varphi^\dagger_+ , i \varphi_-)$
(the  factor $i$ is for notational convenience)
we get, after integration by parts,  the linear Hamiltonian 
\begin{align*}
H_{\rm Lin} = \int dx \Psi^\dag {\cal H}_{\rm Lin}(x)\Psi
\end{align*}
with 
\begin{align}
\label{Hcurl}
{\cal H}_{\rm Lin} = \frac{1}{2}(-iv_F \tau_z \partial_x -
2( {\rm Re}\bigl(\Delta(x)\bigr) \tau_y - {\rm Im}\bigl(\Delta(x) \bigr)\tau_x)) \, ,
\end{align}
where the Pauli matrices now act in right-left spinor-space. The pairing term is  taken so that the
gap for constant $\Delta$ coincides with that in the previous models. 

In the following it will be important that, after rescaling $v_F$ by $1/2$, the Hamiltonian \eqref{Hcurl} is identical to the one used by Takayama, Lin-Liu and Maki (TLM)\cite{Takayama:1980zz}, to describe the zero energy soliton solutions of the polyacetylene chain model introduced by Su, Schrieffer and Heeger (SSH)\cite{Su:1979ut}. We shall therefore refer to it as the TLM model.

A linearized version of the trivial wire described by \eqref{sHam} can be constructed in a similar fashion, but with a four spinor containing the left and right components of the two spin polarizations. For details, see Appendix \ref{sec:AppTop}.

At this point we should point out that the Hamiltonian $H_{\rm Lin}$ presents conceptual 
problems, and does not fit easily into the usual  topological classification. The reasons are as follows:
In Appendix A we show  that  as a consequence of the spectrum in Fig.~\ref{fig:linearize}(c)  
extending from plus to minus infinity, the $k$-space topology is not well defined. Also, the particle-hole symmetry is not a 
consequence of a redundancy due to an artificial doubling of a band. Rather it follows from  extending the linear 
dispersion to arbitrary large negative energies. If we were to add band bending corrections to this model we 
would break the particle-symmetry which again would change the topological classification of the model.  
This situation is unsatisfactory since it raises questions about the validity of linear approximations, and in particular the use of 
the TLM model, for analyzing the Kitaev chain. To resolve this we shall now present an alternative model that 
resolves the problems related to topology and doubling, while retaining a linear spectrum. Having shown the existence of such 
a model, we can safely continue to use $H_{\rm Lin}$ in the subsequent discussion. 


\subsubsection{V-shape model}

First we replace the parabolic band in Fig.~\ref{fig:linearize}(a)
with a V-shaped band, with dispersion
$\epsilon_{\rm v} (k) = | k | v_F - \bar{\mu}$, as shown in Fig.~\ref{fig:linearize}(b).

Next we express the full  field $\psi (k)$ in terms of the low momentum fermion fields $\varphi_\pm (k)$ 
\begin{equation} \label{psi}
\psi (k) = \frac{1}{\sqrt{2}}(e^{i k_F x} \varphi_{+} (k) \vartheta (k) + e^{-i k_F x} \varphi_{-} (k) \vartheta (-k)) \, ,
\end{equation}
where $\vartheta(k)$ is the step function.
In order to write  a BdG Hamiltonian, we first define
\begin{equation} \label{chi}
\chi (k) = \frac{1}{\sqrt{2}}(e^{i k_F x} \varphi_{+} (k) \vartheta (-k) + e^{-i k_F x} \varphi_{-} (k) \vartheta (k)) \, ,
\end{equation}
and  the Nambu spinor $\Phi^\dagger = (\psi^\dagger, -i\chi)$.
Next we substitute \eqref{psi} and \eqref{chi} in the expression for $H_p$, and disregard 
the rapidly oscillating  terms $\sim e^{\pm 2i k_F}$ to get
\begin{equation} \label{vmodel}
\mathcal{H}_{\rm v}(k) = \frac{1}{2}\Phi^\dagger \bigl[ (-\bar{\mu} + v_F |k| ) \tau_z +
\Delta\, {\rm sgn}(k)  \tau_y \bigr] \Phi \, ,
\end{equation}
where again the Pauli matrices $\tau_i$ act in the Nambu space.
As usual, this amounts to a doubling of the spectrum, and this redundancy  is manifested in
the particle-hole symmetry of $\mathcal{H}_{\rm v}$ which cannot be broken. 
The pairing term $\sim \Delta$ (which is assumed to be real)
is  such that it gives rise to the same gap as the original Hamiltonian $H_p$ for constant $\Delta$. 

By inspection, we see that the dispersion relation for $\mathcal{H}_{\rm v}(k) $ has an unphysical $2\Delta$ jump at $k=0$.
This discontinuity can be regularized by smoothening the tip of the V-shaped band, and this will in fact be necessary when we analyze the topological properties
in Appendix A. Such a regularization will however necessarily yield a more complicated model, that is only amenable to numerical solutions, in spite of having a very simple low energy limit. We will not pursue this since, this model is of interest only to demonstrate the existence of a consistent model with a linear spectrum, and 
good topological properties.


\section{Junctions and solitons}
\label{sec:soliton}

\subsection{Topological properties}

We  start our discussion of $\pi$-junctions in 1D superconductors, by reminding
the reader about which different topological superconductors are possible in 1D
systems. To do this, we recall the topological classification of non-interacting fermion systems
\cite{Schnyder2008,KitaevPeriodic,RyuLudwig}, where the possible topological
phases are classified according to their non-unitary symmetries, \emph{viz.}  time-reversal symmetry (TRS)
$\mathcal{T}$ and particle-hole symmetry (PHS) $\mathcal{C}$ (we note that the PHS is technically a spectral constraint rather than a physical symmetry. However, we here chose to follow the widely adopted terminology of \oncite{Schnyder2008}).

In this paper, we consider superconductors in one dimension without spin rotation symmetry. The BdG structure of the Hamiltonian entails a built in algebraic constraint rooted in the fermionic algebra of the field operators that can formally be viewed as a PHS
with $\mathcal{C}^2 = +1$.
In the absence of time-reversal symmetry, \ie for class D, the superconductor is 
either  topologically  trivial, or  non-trivial, depending on the value of the $\mathbb{Z}_2$ invariant. 
In the latter case the wire supports MZMs at both ends \cite{Kitaev2001}.
 In the case of time-reversal symmetric superconductors, with
$\mathcal{T}^2  = -1$, \ie in class DIII, the situation is similar, but in this case, the topological phase
exhibits a Kramers-degenerate pair of MZMs at both edges, see, e.g., Refs. \onlinecite{Law1DD3,Nakosai1DD3}.)
Finally, if the system respects the PTRS $\mathcal{T}^2 = +1$, \ie for class BDI, the different topological phases are distinguished by 
an integer winding number, giving an infinite set of different topological non-trivial phases.

The $p$-wave wire, \eqref{contsc} or \eqref{kitaevstart}, will 
in general, \ie when we allow both the hopping and the order parameter to be complex, 
belong to symmetry class D, which means that it can either be in a trivial phase, or
in a topological phase. In the lattice model, the former happens for $|\mu| > 2 |t|$, while the latter occurs for
$|\mu| < 2 |t|$, with $|\Delta| \neq 0$.

If both $t$ and $\Delta$ are real, the Hamiltonian \eqref{kitaevstart} is also pseudo time-reversal
symmetric (here $\cal T$ is simply complex conjugation, so trivially ${\cal T}^2 =1$), and in this case,
the possible topological phases are labeled by an integer, corresponding to a winding number
(see Appendix~\ref{apptoplin}).
Kitaev's model with a constant order parameter exhibits three of these phases,
namely the trivial one (when $|\mu| > 2 |t|$), as well as two non-trivial ones,
both occurring for $|\mu| < 2 |t|$, one with $\Delta > 0$, the other with $\Delta < 0$.

This means that for real $t$ and $\Delta$, the Kitaev chain can harbor 
an interesting junction, by allowing the order parameter
to change from $-\Delta$ to $+\Delta$ in a finite region, corresponding to a $\pi$-junction. 
Just as the edge of a Majorana wire hosts a MZM, because it constitutes 
the boundary between a topological phase and the trivial vacuum, 
the $\pi$-junction we consider here will also support  zero modes. 
Since the difference in winding number between the two
neighboring topological phases is two, we expect twice as many zero modes
in comparison to the edge of the Kitaev chain. 
Below we show that this is indeed the case, 
irrespective of the precise $x$-dependence of the order parameter.

We already mentioned the problems related to properly define the $k$-space topology
for the TLM model, and how they are resolved by an alternative linearization scheme. 
The details are given in Appendix A, but we should here again stress that the outcome
of this analysis is that we can safely use the TLM model to discuss the topological
properties of the Kitaev chain.

\subsection{The $\pi$-junction as a soliton}

Although the $k$ space argument for topology of the linearized model ${\cal H}_{\rm Lin}$
given in Appendix \ref{apptoplin} is compelling, it is important to find out how well the TLM model \eqref{Hcurl} really captures the topological properties of the full model \eqref{kitaevstart}.
To do this, we first briefly recall how a Dirac Zero Mode (DZM) arises in the TLM
model\cite{Takayama:1980zz}, and then compare it with the zero mode arising in the full model
\eqref{kitaevstart}, in the presence of a junction, at which the real order parameter $\Delta$
changes sign.

The presence of the DZM in the case of a real order parameter is most easily demonstrated in
the TLM model, and from  ${\cal H}_{\rm Lin} $  we get the BdG equations
\begin{align}
\frac{1}{2}(-i v_F \partial_x u(x) + 2 i  \Delta^* (x) v^*(x)) &= \epsilon u(x) \nonumber \\
\frac{1}{2}( i v_F \partial_x v^*(x) - 2 i  \Delta (x) u(x)) &= \epsilon v^*(x) \ .
\label{eq:bdglin}
\end{align}
For  real $\Delta(x)$, and taking $\epsilon = 0$  since we are interested in the zero modes,
these equations are easily decoupled by introducing $f_\pm  (x)= u (x) \pm v^* (x)$.
For a $\pi$-junction that interpolates between a negative constant
$\Delta_-$ for $x\ll 0$ to a positive constant $\Delta_+$ for $x \gg 0$, 
one finds the solution
$f_+(x) = N e^{-\frac{2 k_F}{v_F} \int^x \Delta (x') dx'} \ , f_-(x) =0$.
Here, we will consider the special profile
$\Delta(x) = \Delta_0 \tanh(x/\xi)$, that gives rise to the analytical solution\cite{Takayama:1980zz}
\begin{align}
f_+(x) &= N_0\,  \text{sech}(x/\xi)^{\xi/\xi_0} & f_{-} (x) &= 0 \ ,
\end{align}
where $\xi_0 \equiv v_F/(2\Delta_0)$ and $N_0$ a normalization constant.

\begin{center}
\begin{figure}[hbt]
\includegraphics[scale=0.65]{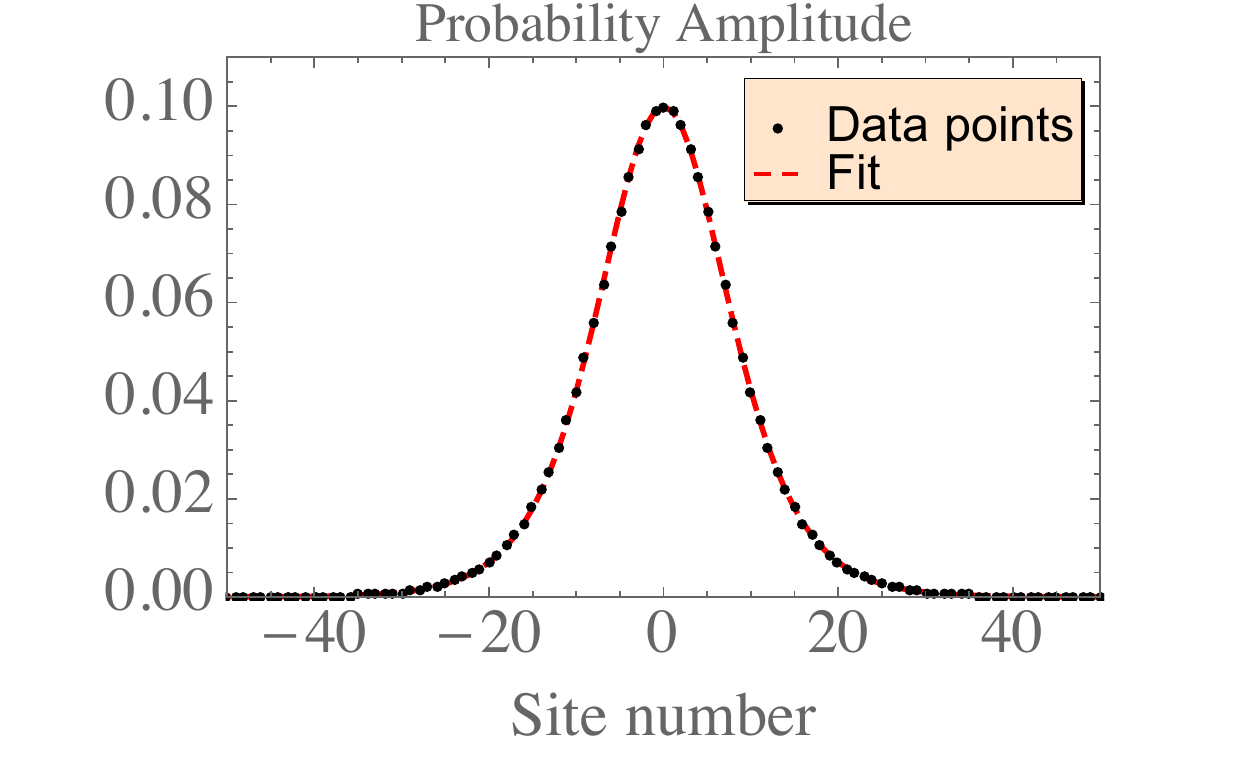}
\caption{The probability distribution (black dots) of the zero energy mode located in the junction region of the Kitaev chain with order parameter profile $\sim \tanh(x/\xi_0)$. The parameters used are: $t = 10.0$, $\Delta_0 = 1.0$ and $\mu = 0.0$ yielding $\xi_0 =10$. The fit (red, dashed line) is made by the method of least squares, and resulted in $\lambda = 9.994$, in good agreement with the value of
$\xi_0$. The number of sites is $N=1001$; the figure only shows the central region.}
\label{fig:zeromode} 
\end{figure}
\end{center}

We compare the DZM of the TLM model to the full model, by considering the discretized version
$\Delta_j=\Delta_0\tanh(j/\xi)$ of the TLM profile $\Delta(x) = \Delta_0 \tanh(x/\xi)$ in \eqref{kitaevstart}.
By choosing the width $\xi$ not too large, this determines the order parameter to $-\Delta_0$ at one end of the chain and $+\Delta_0$ at the other end, generating a domain wall (between sectors with different winding numbers) at the center of the chain. We set the junction parameter $\xi=\xi_0$, and
fit the resulting zero mode to the TLM solution $g(x) = {\rm sech} (x/\lambda)$, with $\lambda$ used as
a fitting parameter. In Tab. \ref{Tab:easyfits} we show some representative results and in Fig.~\ref{fig:zeromode}, we display a typical result for the probability density of the DZM located in the junction, as obtained from the Kitaev chain. Evidently, the TLM model captures the the properties of the DZM in the junction region of the Kitaev chain very well.

\begin{table}
  \begin{tabular}{c | c | c | c | c  }
    $t$ & $\Delta_0$  & $\mu$ & $\lambda/\xi_0$ & MLS Error \\ \hline
    $10.0$ & $1.0$ & $0.0$ & $0.999418$ & $2.2002\cdot10^{-8}$ \\ 
    $8.0$ & $1.0$ & $0.0$ & $0.999098$ & $6.7994\cdot10^{-8}$ \\ 
    $5.0$ & $1.0$ & $0.0$ & $0.997777$ & $7.5333\cdot10^{-7}$ \\ 
    $2.0$ & $1.0$ & $0.0$ & $0.991853$ & $1.2267\cdot10^{-4}$\\
    $1.0$ & $1.0$ & $0.0$ & $1.071428$ & $6.8020\cdot10^{-3}$\\
  \end{tabular}
\caption{
The fitting parameter $\lambda$, compared to its analytic value in the TLM model $\xi_0$, as well
as the error of the fit (using the Method of Least Squares), for a system with
$N=1001$ sites and various values of $t$.}
\label{Tab:easyfits}
\end{table}


\section{Phase winding junctions} 
\label{sec:phasewindingjunctions}
In this section, we extend the previous discussion to the case of junctions with a complex order parameter where  the phase winds in a finite segment of the wire. For simplicity we shall assume that the absolute value  $|\Delta|$ remains constant. In this case we can find an analytical solution in the linearized model by taking a simple winding profile, while our numerical analysis easily generalizes to more general profiles.

Although a complex order parameter breaks the PTRS and with that the chiral symmetry that protects the DZMs in the junctions with real profiles, one would still expect that the low energy theory should not  differentiate a rapid winding of the phase from 0 to $\pi$ from a sharp step in the magnitude of $\Delta$.   We now demonstrate that this intuition is correct, and that low energy modes persist even in the case of phase winding junctions. Again it is easiest to start from the linearized model. 

\subsection{sub-gap states in the linearized model} 
\label{sec:sub-gap}
We consider an order parameter with an $x$-dependent phase
\begin{align}
\Delta &= \Delta_0 e^{i \theta(x)} &
\theta (-\infty) &=0 &
\theta (\infty) &=f\pi,
\label{solprof}
\end{align}
where $\Delta_0$ is a positive constant, $\theta(x)$ is continuous and $f$ is some real number. The BdG equations \eqref{eq:bdglin}
then become,
\be{zerom}
- i \partial_x u(x) +i /\xi_0 e^{- i \theta (x)} v^* (x) &= \tilde{\epsilon} u(x) \\
i \partial_x v^*(x) - i /\xi_0 e^{i \theta (x)} u(x) &= \tilde{\epsilon} v^*(x) \nonumber \, ,
\end{align}
where $\xi_0 = v_F/(2 \Delta_0)$ and $\tilde{\epsilon} = 2\epsilon/v_F$.
From the first equation we have
\be{solve}
v^*(x) = -i \xi_0 e^{i \theta (x)} (\tilde{\epsilon} + i \partial_x ) u(x) 
\end{align}
and substituting this into the second, we get
\be{secondor}
\left[  \partial_x^2  + i (\partial_x \theta) \partial_x + 
\tilde{\epsilon} (\partial_x \theta)   + (\tilde{\epsilon}^2 - \xi_0^{-2}) \right ]u(x) = 0 \ .   
\end{align}
This equation cannot be solved analytically for a general profile, but for the case of 
\be{specprof}
\theta_k (x) = \left\{ \begin{array}{cc}  0 & x <  -a \\ 
 \left(\frac {x+a} {2a}\right) f \pi &|x| \leq a \\ 
 f\pi & x > a  
 \end{array}\right .
\end{align}
we can solve \pref{secondor} in the three regions and then match the solutions.
Just as in an 1D Schr\"{o}dinger problem in a piece-wise constant potential,
this is done by matching the function and its (logarithmic) derivative.
We focus on the case  $f=1$, which corresponds to a $\pi$-junction where
$\Delta$ changes sign, but the analysis below can easily be extended to junctions with
arbitrary phase winding.

The piecewise solutions are given by
\be{bssol}
u (x) = \left\{ \begin{array}{cc}  \alpha_1 e^{\kappa x}  & x < -a \\  
 e^{-i \frac \pi {4a} x} \left(   \alpha_{2}^+e^{\tilde\kappa x } + \alpha_{2}^-e^{-\tilde\kappa x } 
 \right)  & |x|\leq a \\  
\alpha_{3} e^{-\kappa x}    &  x>a  
\end{array}\right .
\end{align}
where $\kappa = \sqrt{\xi_0^{-2} - \tilde{\epsilon}^2}$
and
$\tilde\kappa = \sqrt{\xi_0^{-2} - \left(\tilde{\epsilon} + \pi /(4a)\right)^2}$.
To obtain a normalizable solution, we must take $\kappa < 0$, or
$|\tilde{\epsilon} | < 1/\xi_0$, implying that the (sub-gap) solution is localized in the
junction region. From the matching conditions for the wave function and its
derivative, one can infer that there is no solution when $\tilde{\kappa}$ is real. An imaginary
$\tilde{\kappa}$ requires that $\tilde{\epsilon} > 1/\xi_0 -\pi/(4a)$, so localized sub-gap
modes are possible in the energy range $1/\xi_0 -\pi/(4a) < \tilde{\epsilon} < 1/\xi_0$
if $a > \xi_0 \pi/8$, or in the whole gap region $- 1/\xi_0 < \tilde{\epsilon} < 1/\xi_0$
if $a < \xi_0 \pi/8$.

For imaginary $\tilde{\kappa}$, the matching conditions have a solution if
the following constraint is satisfied
\begin{equation}
\begin{split}
\label{numEns}
\tan \Bigl( 2a \sqrt{ (\tilde{\epsilon} + \frac{\pi}{4a})^2 - \xi_0^{-2}} \Bigr)
=\\
\frac{\sqrt{\xi_0^{-2}-\tilde{\epsilon}^2}\sqrt{(\tilde{\epsilon} + \frac{\pi}{4a})^2 - \xi_0^{-2}}}
{\tilde{\epsilon}^2 + \frac{\tilde{\epsilon}\pi}{4a} -\xi_0^{-2}}.
\end{split}
\end{equation}
Upon analyzing this equation, one finds that even for arbitrary small $a$, there is
always at least one solution. The energy of the associated bound state is always
positive, but approaches zero in the limit of small $a$. Upon increasing $a$, more
and more bound state solutions appear. In order to have at least $p+1$ bound states,
$a$ should satisfy $a \geq \frac{(4p^2-1)\pi\xi_0}{8}$.

Before turning to the numerical results, we briefly discuss the case of general phase
winding, \emph{i.e.}, we allow $f$ in \eqref{specprof} to be arbitrary.
For $f$ arbitrary small, one finds a bound state, with an energy slightly below the band gap,
$\tilde{\epsilon} \lesssim 1/\xi_0$. Upon increasing $f$, the energy of this bound state
decreases towards $\tilde{\epsilon} = -1/\xi_0$. In the mean time, more bound states
appear  at the gap edge $\tilde{\epsilon} = 1/\xi_0$. In the limit of large $f$, the
energies of the bound states become periodic in $f$, with a period of $2$, \emph{i.e.}, a period
of $2\pi$ in the winding angle. Finally, in the limit of a very short junction, we find that
for $f$ an odd integer, there is a bound state at $\tilde{\epsilon} \approx 0$, while for
$f$ an even integer, there are two bound states with energy $\tilde{\epsilon} \approx \pm 1/\xi_0$.
In the former case, the junction behaves as a $\pi$ junction with a real order parameter,
while the second case is equivalent to not having a junction at all. This is consistent with
the topological discussion  above, although we should point out
that there are no topological reasons why the phase junction should behave as a real
junction in the short junction limit. We next compare some of the results of this section with 
numerical simulations in the Kitaev chain and in the full $s$-wave model. 

\subsection{Comparison with the Kitaev chain}
Starting with the Kitaev chain model given in \eqref{kitaevstart}, we take the profile $\Delta_j = |\Delta_0| e^{i\pi\frac{(j+a)}{2a}}$,  so that over a segment of length  $2a$, the phase increases linearly  from $0$ to $\pi$. Effectively, this amounts to changing the sign of $\Delta$ just as in the previous section. 
Using this profile, we numerically calculated the energy of the low lying fermion states both for the Kitaev chain and the
linearized model, using a range of parameters. Typical results are shown in Tab. \ref{Tab:phasestates}, where the agreement between the first two columns is a measure of the precision of our numerics, and the good agreement with the third column again confirms that the linearized model faithfully describes the full Kitaev chain. We have also compared the numerical wave functions for the low lying states in the Kitaev chain, with the analytical expressions \eqref{bssol} and again found excellent agreement.

\begin{center}
\begin{figure}[ht]
\includegraphics[scale=0.65]{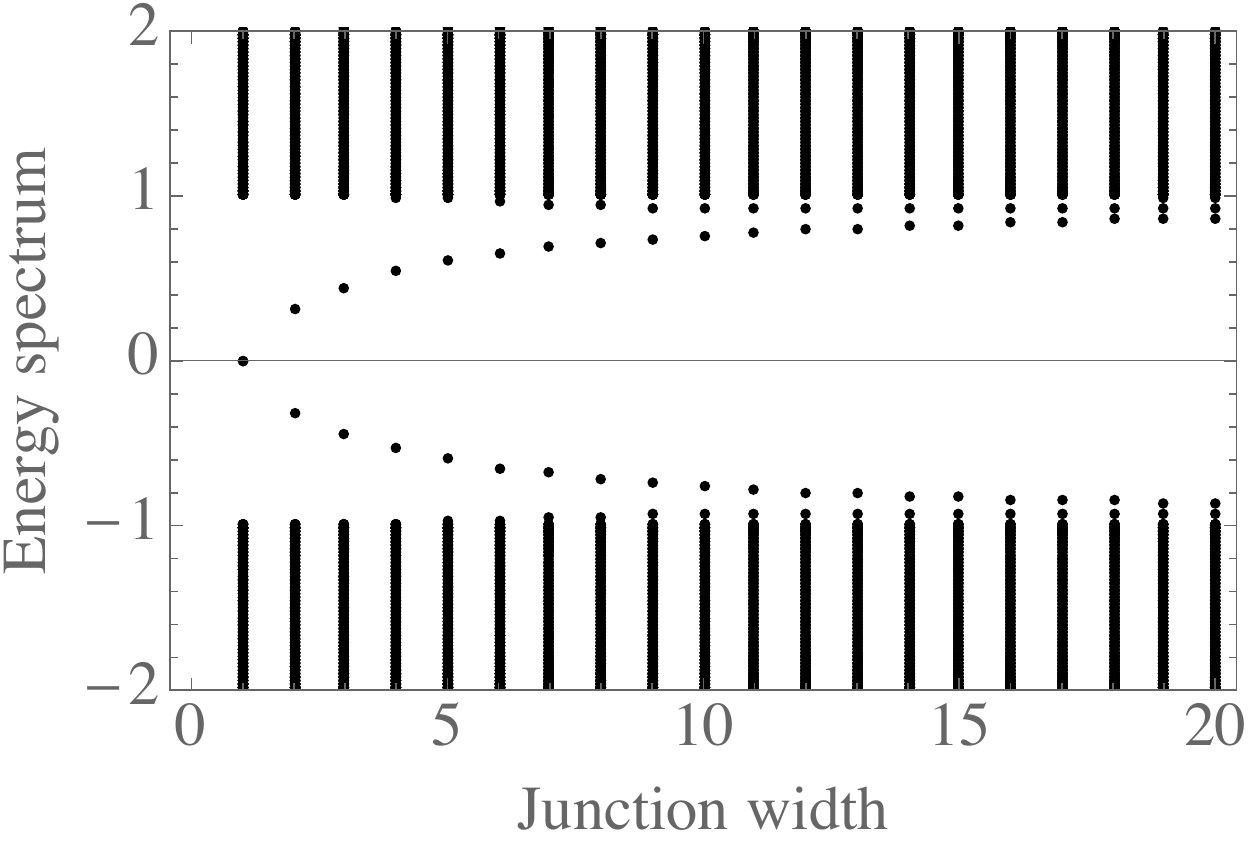}
\caption{The energy spectrum of the Kitaev chain as a function of its $\pi$ phase junction length (in units of the lattice parameter).. A junction length of $1$ means that the phase jumps from $0$ to $\pi$ from one site to another. Note that the zero energy states that represent MZMs located at the end points of the chain have been omitted. In addition, two new zero modes are formed as the junction length shrinks, effectively imitating a real $\pi$-junction. The parameters used are $t=2.0$, $\Delta_0 = 1.0$, $\mu=0.0$ and $N=200$. The spectrum is displayed in a low energy regime.}
\label{fig:PWindingJunction} 
\end{figure}
\end{center}

\begin{center}
\begin{figure}[ht]
\includegraphics[scale=0.65]{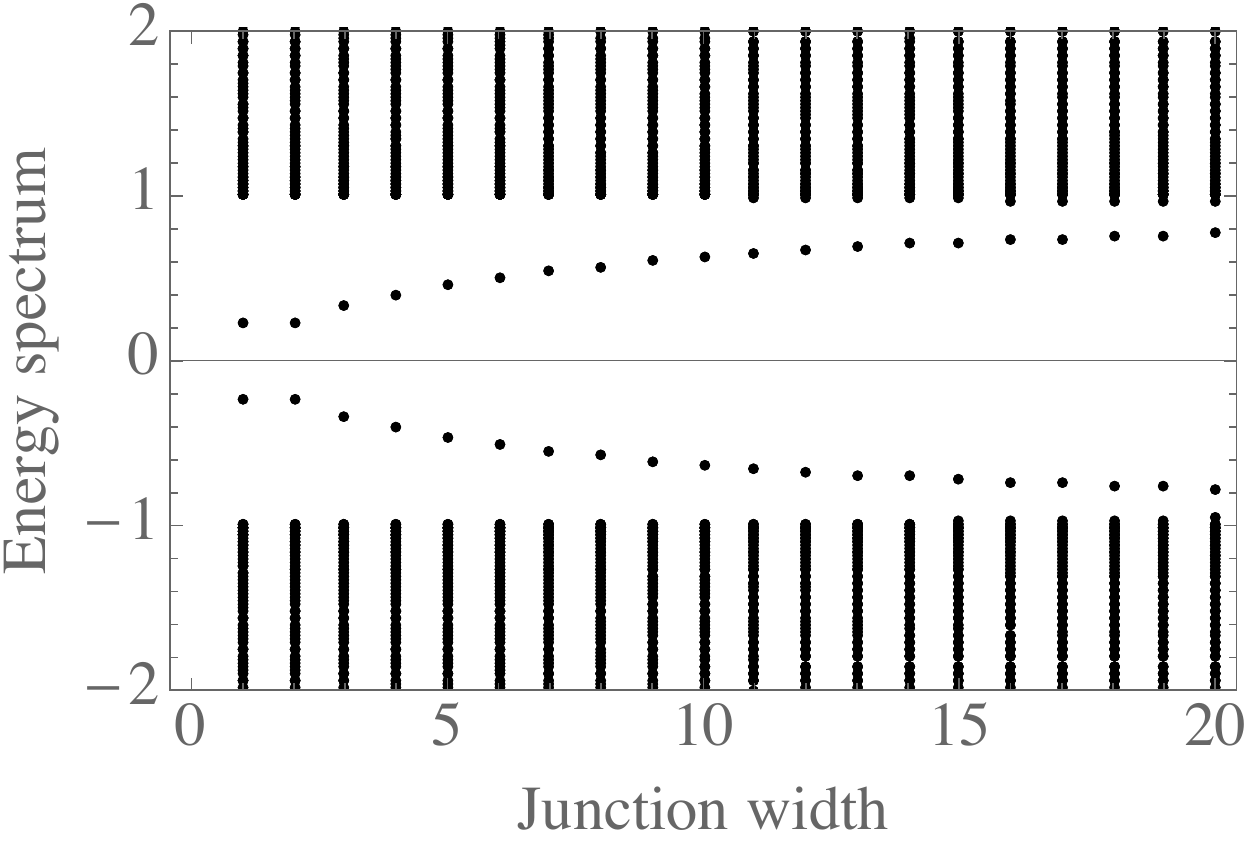}
\caption{The energy spectrum of the $s$-wave wire as a function of the length of its $\pi$ phase junction (in units of the lattice parameter). A junction length of $1$ means that the phase jumps from $0$ to $\pi$ from one site to another. Because the $s$-wave wire is topologically trivial, no zero modes form, even in the limit of a short junction. The parameters used are $t=2.0$, $\Delta_0 = 1.0$, $\mu=0.0$ and $N=200$. The spectrum is displayed in a low energy regime.}
\label{fig:SWindingJunction} 
\end{figure}
\end{center}

Next we studied what happens when the length of the phase winding $\pi$-junction shrinks. In Fig. \ref{fig:PWindingJunction}, which shows our result for the $p$-wave case, we see clearly how a state that is close to the gap for large junctions comes down, and becomes a zero mode for the shortest junctions (which essentially amounts to a sign change between two lattice points). This supports the heuristic argument, given earlier, that a short phase winding $\pi$-junction should have properties very similar to the one where $\Delta$ remains real but changes sign. The corresponding $s$-wave setup is depicted in Fig.\ref{fig:SWindingJunction}, where no zero modes need to be formed in the short junction limit. 

These results give additional confirmation that the low energy properties of junctions made by Kitaev chains can be captured by the linearized model in \eqref{linsc}, and in the Section  \ref{sec:effectivetheory} we construct a topological field theory, which captures the same physics.

\begin{table}
  \begin{tabular}{c | c | c }
    Analytic sol. & Linear model & Full model  \\ \hline
    0.940199 & 0.940201  &0.940156  \\
    0.956549 & 0.956556 & 0.956385 \\
    0.981316 & 0.981324 & 0.981002 \\    
  \end{tabular}
\caption{The energies of the first three bound states in a $p$-wave $\pi$ phase winding junction. The parameters used for these calculations are: $t=10.0$, $\Delta = 1.0$, $\mu=0.0$, $a=120$, $N=800$. Analytical, linear model and full model refer to the equations \eqref{numEns}, \eqref{eq:bdglin} and \eqref{kitaevstart} respectively. Note that the linear model values are just a measure of how well analytic solution describes the discretized linear model,  
while the full model values describe how well the linearization captures the low energy degrees of freedom.}
\label{Tab:phasestates}
\end{table}

\section{How to experimentally probe topology by a $\boldsymbol{\pi}$-junction \label{sec:exp}}

Most of the experimental effort in studying the topological wires has been aimed at detecting the MZM at the edges.
But as mentioned in the introduction, the proposed signatures for these modes can also be emulated by other effects.
It is thus interesting to consider other signatures for the wire being in the topological phase, and here we suggest the possibility of using the DZMs at   $\pi$-junctions as  such a probe. For this idea to be useful, we not only need a way to experimentally realize such a junction and detect the associated fermionic zero modes, but also a clear signature for the topological phase. We shall consider both junctions with topologically protected zero modes, and phase winding junctions. We begin with the latter.

\subsection{Phase winding \pij}

One way to make a phase winding junction is to put a  wire of the type used in previous experiments on top of a 
$s$-wave superconductor through which a current is driven between external leads placed close to the wire. 
By the the relation $\triangledown \phi \sim J$, where $\phi$ is the superconducting phase and $J$ is the current, one can arrange for a $\pi$ phase difference between the leads, which will, by proximity, be imprinted on the wire. An experimentally more challenging task is to probe the fermion spectrum at the junction. An obvious possibility is to use a tunneling contact weakly coupled to the wire, or a scanning tunneling microscope.

From the previous section it would appear that a good signature for the $p$-wave pairing phase would be the presence of an almost zero mode in the junction region. Unfortunately, the situation is not very clear since an $s$-wave pairing would have a similar signature. Fig. \ref{fig:SWindingJunction} is similar to  Fig. \ref{fig:PWindingJunction}, but for $s$-wave pairing. Also here we find a low-lying sub-gap state for short junctions, and although it does not come all the way to zero, it is not clear that it could be distinguished from the $p$-wave case. Clearly one would need much more detailed studies of more realistic microscopic models in order to resolve this question. 

\subsection{Real \pij}
\begin{center}
\begin{figure}[ht]
\includegraphics[scale=0.65]{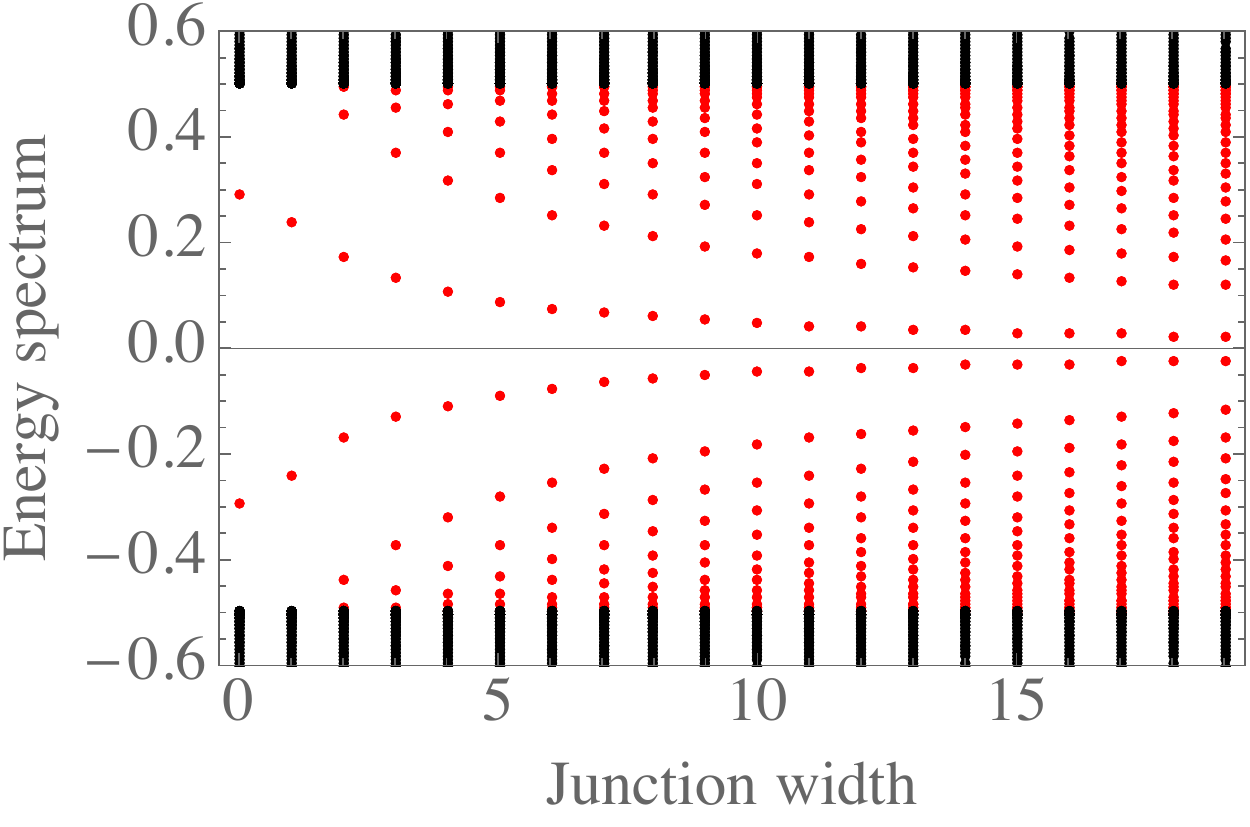}
\caption{The energy spectrum (bulk states in black and subgap states in red) of the $s$-wave wire (a discrete version of Eq. \eqref{sHam}) as a function of the width of its real $\pi$-junction (in units of the lattice parameter). Modes with zero energy exist only in the limit of a wide junction, and are gapped out in the short junction region due to the low chemical potential. The parameters used are $t=1.0$, $\Delta_0 = 1.0$, $\mu=1.9$ ($\bar{\mu} = 0.1$) and $N=200$. The spectrum is displayed in a low energy regime.}
\label{fig:SReal1} 
\end{figure}
\end{center}

\begin{center}
\begin{figure}[ht]
\includegraphics[scale=0.65]{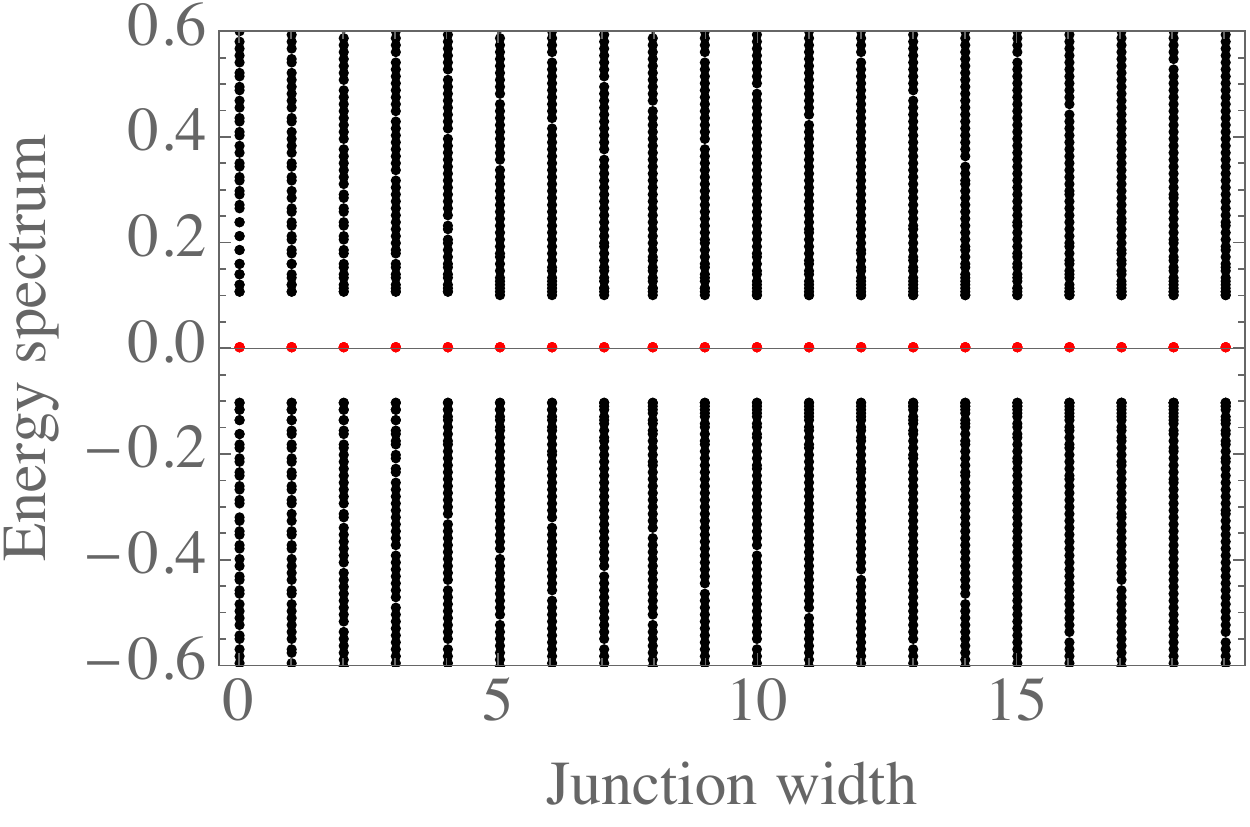}
\caption{The energy spectrum (bulk states in black and subgap states in red) of the $p$-wave wire (Eq. \eqref{kitaevstart}) as a function of the width of its real $\pi$-junction (in units of the lattice parameter). The Dirac zero mode is topologically protected and exists for short and long junctions since the wire is in its topological phase. Subgap modes with finite energy are not present in the junction length regime displayed here. The parameters used are $t=1.0$, $\Delta_0 = 1.0$, $\mu=1.9$ ($\bar{\mu} = 0.1$) and $N=200$. The spectrum is displayed in a low energy regime.}
\label{fig:PReal1} 
\end{figure}
\end{center}

As already pointed out, in a \pij with a real order parameter (that must go through zero) the zero energy Dirac mode is always present when the superconductor is in the topological phase. For the trivial $s$-wave case, there is no such protected zero mode, but the spectrum of the subgap modes does depend on the profile of the order parameter at the junction (and on the other parameters, such as the chemical potential). Importantly, there can be junction modes with zero energy, that can be described by the TLM model we studied above, for certain choices of parameters.
For example, putting the chemical potential in the band middle ($\mu = 0$ or equivalently $\bar{\mu} = 2t$ as measured from the bottom of the band), there are localized modes with zero energy, regardless of the junction length. But these states can be gapped out in the short junction limit by lowering the chemical potential to the vicinity of the band bottom. This feature is demonstrated in Fig.~\ref{fig:SReal1} and contrasted with the corresponding $p$-wave system in Fig.~\ref{fig:PReal1}. In the latter case, the topology protects the zero mode, regardless of the junction length, as long as the chemical potential lies in the band, so that the system is in the topological phase. 

We note that for wide $p$-wave junctions, $\xi \gtrsim 30$, there are additional subgap modes with finite energy which are not in the range of $\xi$-values in Fig.~\ref{fig:PReal1}.

There are at least two possible ways to experimentally realize a junction of this type. The most direct would be to implement a \pij in the underlying $s$-wave superconductor, but the problem here is that it is not easy to determine the $\Delta$ profile in the junction. An alternative way is to put the wire as a bridge connecting two different $s$-wave superconductors that are held at different values of $\theta$, for instance by a SQUID geometry. In this case one might calculate the $\Delta$ profile by a realistic modeling of the wire.

Clearly both options need to be studied in more detail before any definite conclusion can be made about the feasibility of using sub-gap junction spectroscopy for probing the topological nature of the wire.


\section{A topological field theory for $\boldsymbol{\pi}$ phase winding junctions}
\label{sec:effectivetheory}

As in the previous sections we consider $\Delta$ as given by the background
$s$-wave superconductor by proximity. 
We generalize the previous discussion somewhat by considering a wire with many widely
separated $\pi$-junctions of the phase winding type. 
Since the bound states are exponentially localized, such configurations will support  subgap modes at each junction.
 We stress that we consider a fixed junction configuration, given by the background 
$s$-wave order parameter, and derive an effective theory for the fermions. We can, using the same formalism, also 
describe adiabatic motion of the junctions, but they cannot be considered as  {\em bona fide} itinerant particles.

Our  starting point  is the Lagrangian formulation of
the linearized version of the $p$-wave superconductor \eqref{Hcurl},
which is given by
\begin{align}\label{diraclag}
{\cal L} = \psib \left( i \slashed \partial - g(x)e^{i\vartheta (x)\gfem } \right) \psi \ ,
\end{align}
where the functions $\vth$ and $g$ are related to the superconducting order parameter by
$\Delta = g e^{i(\vth +\pi/2)}$, and where the Dirac matrices are related to the Pauli matrices by
$(\gamma_0,\gamma_x ) =(\sigma^x, -i\sigma^y)$ so $\gfem = \sigma^z$. We have in our derivation relabeled the spinor $\Psi$ to $\psi$, in accordance with standard notation. Furthermore, we have put $v_F =2$, consistent with the linearization of \eqref{contsc}. Note that for real $\Delta$, the energy gap $\Delta$ is nothing but the mass in the Dirac equation. 

The aim here is first to derive a bosonic form of the Lagrangian \eqref{diraclag}, and then to extract an effective action that describes the physics of the bound states on the solitons.  This theory is topological in the sense that it does not have any bulk degrees of freedom, but only describes the quantum mechanics of the bound states residing on the solitons. Before embarking on this exercise, we will put it in context, and view it as part of the more challenging problem to formulate topological theories in the presence of fermionic zero modes. 

As mentioned in the introduction, the Kitaev chain is a one-dimensional cousin of the two-dimensional (2D) $p$-wave superconductor, and the Majorana states located on the interfaces between the normal and topological phase of the model can, by employing geometries with junctions, be used for quantum computing. In this context the quantum mechanics of the zero modes is clearly very interesting. In the 2D case, there are two candidates for a topological field theory that describes the braiding of vortices with Majorana zero modes. One is based on a $SU(2)$ Chern-Simons theory\cite{fradkin}, while the other employs an abelian BF theory coupled to a single Majorana field\cite{hansson12}. In this 2D case, the vortices are in principle itinerant, but are in practice often pinned to impurities. In this latter case there is a close analogy with our system of fixed, or adiabatically moving, $\pi$-junctions.

None of these effective theories just mentioned has been derived from a microscopic description, but are obtained from general principles based on symmetry and scaling. The effective topological theory for the fermionic bound states on solitons that we shall describe shortly, is closely related to the second of the 2D topological theories that we just mentioned. An obvious, and important, difference is that the fermionic modes on the solitons are of Dirac type, so, even if fine tuned to zero energy, they  can  not be used for topological quantum computing. The advantage with the present case is that it is more amenable to analytical treatment. Still we have not managed to obtain the topological theory directly from the microscopic model in a controlled fashion. The derivation  presented below is therefore phenomenological and again based on symmetry considerations and scaling arguments. In  Appendix \ref{efftheoryapp} we do offer a microscopic derivation which however involves several unproven, and admittedly questionable, assumptions. 
 
 \subsection{Symmetries}
We now discuss the symmetries of \eqref{diraclag}. From this Lagrangian we  can immediately get the vector and axial charge densities,
\begin{eqnarray}
\rho_V &=& \psid \psi = \vph_+^\dagger \vph_+ -  \vph_-^\dagger \vph_-   \label{vcharge}\\
\rho_A &=&\psid \gfem \psi = \vph_+^\dagger \vph_+ + \vph_-^\dagger \vph_-  \label{acharge}
\end{eqnarray}
which shows that the electric charge $Q_{em} \equiv Q_A = \int dx\, \rho_A(x)$ in the superconductor is given by the {\em axial} charge \pref{acharge} in the  Dirac theory \eqref{diraclag}, and is thus not conserved, as appropriate for a superconductor. Note, however that \eqref{diraclag} {\em is} invariant under the
combined global transformation
\be{emsymm}
\psi &\rightarrow e^{i\beta \gfem} \psi \\
\theta &\rightarrow \theta - 2\beta \nonumber \, .
\end{align}
In a BdG description this corresponds to a simultaneous global phase change of the electron field and the superconducting condensate $\av{\psi\psi}$.
Also note  that the transformation,
\be{dsymm}
\psi \rightarrow e^{i\pi\gfem} \psi = - \psi
\end{align}
is indeed a symmetry. As expected, this is a manifestation of the  the conservation of electric charge modulo two, which is most easily seen by noting that the transformation \pref{dsymm} leaves the pairing terms $\vph_+^\dagger\vph_-^\dagger$ and $\vph_+\vph_-$ invariant. It will be important later that the vector charge $Q_{V} = \int dx\, \rho_V(x)$ in the Dirac theory is indeed conserved. Physically this is a consequence of the Cooper pairs having zero momentum, so adding or subtracting a pair will give identical changes at the two Fermi points\cite{stone-gaitan}. In the following we shall give a bosonized version of the theory where it will be important to keep the correct symmetry pattern.

\subsection{Bosonization}
It will be advantageous to rewrite \pref{diraclag} in bosonic variables using the bosonization  ``translation table'', (see for instance \oncite{fradkin-ftcmp}),
\begin{align}
\psib \psi &\rightarrow \kappa \cos \varphi  \nonumber \\
\psib\, i\gfem \psi &\rightarrow \kappa \sin \varphi    \\
\psib \gamma_\mu \psi &\rightarrow \frac 1 {2\pi} \epsilon^{\mu\nu} \partial_\nu \varphi \nonumber
\end{align} 
where the dimension-full parameter $\kappa$ depends on the short distance cutoff, and the scalar field $\varphi$ is normalized so that the bosonic version of
\eqref{diraclag} is
\be{blag}
{\cal L} = \frac 1 {8\pi} (\partial_\mu\varphi)^2 - g \cos (\varphi - \vartheta) 
\end{align}
where we have rescaled $g$ with $\kappa$. The minima of the potential are at
\be{winding} 
\varphi_n =  \vartheta + \pi + n 2\pi = \theta - \frac \pi 2 + n 2\pi \, ,
\end{align}
so for large $g$, $\vartheta = \theta - \pi/2$ will make small fluctuations around one of these (equivalent) minima. In particular, if $\theta$ winds, then $\varphi$ follows. 
From the work of Jackiw and Rebbi\cite{jackrabbit}, and Goldstone and Wilczek\cite{goodwill}, we know that windings in the scalar field $\varphi$ will describe solitons carrying (in general fractional) fermion number. 
For simplicity we  neglect $2\pi$ windings, and taking $n= 0$ in \pref{winding} we define the kink current as
\be{kinkcurr}
j^k_\mu = \frac 1 {2\pi} \epsilon^{\mu\nu} \partial_\nu \theta \ , 
\end{align}    
so the charge of the soliton that interpolates $\theta(x)$ from $\varphi_L$ to  $\varphi_R$ is given by,
\be{solch}
Q_s = \frac 1 {2\pi} (\varphi_\mathrm{R} - \varphi_\mathrm {L} ) \ .
\end{align}
It follows that  the $\pi$-junctions we discussed earlier carry a half unit of fermion number.

Next we shift the field $\varphi$ by $\varphi = \varphi_0 + \phi$ to get
\be{shiftlag}
{\cal L} &= \frac 1 {2\pi} \epsilon^{\mu\nu} \partial_\nu\theta \, b_\mu - b_\mu j^\mu_k \\
&+ \frac 1 {8\pi} (\partial_\mu\phi)^2  - g \cos (\phi)  + \frac 1   2j_k^\mu \epsilon_{\mu\nu} \partial^\nu \phi \nonumber 
 + \frac 1 {8\pi} (\partial_\mu\theta)^2
\end{align}
where $b_\mu$ is a multiplier field that imposes the condition \pref{kinkcurr}.
Since the $\phi$-field  is massive, it can be integrated, to yield the truly trivial topological theory,
\be{toplag}
{\cal L}_{\theta b} = \frac 1 {2\pi} \epsilon^{\mu\nu} \partial_\nu\theta \, b_\mu - b_\mu j^\mu_k.
\end{align}

\subsection{Retaining the fermion bound states}
The topological theory we just derived is however not always a good description of the low energy physics. This is most easily seen by considering the special case where the topological current describes widely separated narrow $\pi$-solitons. 
We learned in section \ref{sec:sub-gap} that these can support low energy fermionic bound states with energy $\epsilon_0<\epsilon<\Delta$ inside the gap. Since we furthermore can fine tune so one of these modes occurs arbitrarily close to zero energy,  the topological theory \pref{toplag} can  clearly not  be universally correct. Moving away from the $\epsilon_0$ point, but still having the bound state far below the bulk gap, \ie $\epsilon_0 \ll g$, it would still be desirable to have a theory that describe these low-lying excitations. What went wrong in the derivation of \pref{toplag} is that while the bosonic fluctuations with energy $\ge g$ were integrated, the more important fluctuations changing the fermion number were not taken into account. We will now remedy this and present a  model that properly includes the dynamics of the low-lying fermionic bound states. 

We shall first  construct a model in the limit of widely separated point-like kinks. Any real  function $\Delta$ that interpolates between $\pm|\Delta_0|$ at $x = \pm\infty$ supports a zero mode. The  kink,  $ |\Delta| \eta(x)$, where $\eta$ is the step function, can be thought of as a limit of such functions, and thus supports a zero mode. 
Also, as discussed above, we  get an approximate zero mode for constant $|\Delta | = m$, and a rapid winding of the phase $\theta$ an odd number of $\pi$. 
In both these cases the topological current related to the kink can be be described by 
\be{singcurr}
j_k(x,t) = \sum_{a=1}^N \delta (x - x_a) \left( 1, \dot x_a \right)
\end{align}
where we allowed for the kink at position $x_a$ to move with velocity $\dot x_a$. 

It is now straightforward to write a Lagrangian for the bound states residing on the kinks,
\be{singlag}
L &= \sum_{a=1}^N \bxi _ai \frac d {dt} \xi_a  \\ 
&= \sum_{a=1}^N \bxi_a(t, x_a(t)) i(\partial_t - \dot x_a(t)\partial_x) \xi_a(t, x_a(t))  \nonumber \\
&= \int dx\, j_k^\mu\,  \bxi (x,t) i\partial_\mu \xi (x,t)  \, .  \nonumber
\end{align}
Combining this with the term \pref{toplag}, yields 
\be{xithblag}
{\cal L}_{\xi \theta b} = \frac 1 {2\pi} \epsilon^{\mu\nu} \partial_\nu\theta \, \left[ b_\mu  + \bxi i\partial_\mu \xi\right] -\epsilon_0\bxi\xi    - b_\mu j^\mu_k
\end{align}
where we also introduced a chemical potential $\epsilon_0$ that fixes the energy of the bound state. 
We shall take $\xi$ to be a complex fermionic field  (otherwise it would not describe a single bound state), but note that it differs from a conventional Dirac fermion in being dimensionless. 

The first term in Lagrangian \pref{xithblag} is closely related to the topological Lagrangian  for a spin-less 2D chiral superconductor given in Ref.~\onlinecite{hansson12}. The main difference is that in the 2D case the Dirac fermion $\xi (x,t) $ is replaced by a  Majorana field $\gamma (x,t)$. In the present setting, that would be appropriate for a domain wall between a trivial and non-trivial phase of the wire. The second term $\sim \epsilon_0$ is not topological and is present only for a complex field.
Note that the kinetic term $\sim \bxi\partial_0 \xi$ in \pref{xithblag} has support only where the topological charge does not vanish, and thus there are no bulk degrees of freedom. The above analysis is, however, valid only for point like sources. The generalization to extended sources, that is  the finite size kinks considered in the previous sections, is our next task.

\subsection{Fermion bound states in extended kinks}
Since for a static kink, the  Hamiltonian  in \pref{xithblag} is only a chemical potential, it can  not describe the fermion modes on an extended kink, but instead gives a continuum of states at energy $\epsilon_0$. To get a realistic low energy theory we must thus introduce more terms in the effective Hamiltonian.  Following the usual logic of 
effective theories we shall retain the lowest derivative terms that ensure the correct symmetries. The crucial symmetry here is the broken global $U(1)$ symmetry related to the electric charge. In the linearized theory \pref{diraclag} this is the (global) chiral symmetry \pref{emsymm}. Clearly terms like
$\bxi \xi$, $\bxi \partial_x^2 \xi$ \etck are allowed, but also pairing terms like $e^{i\theta} \bxi \partial_x \bxi$ \etcp In fact it is necessary to include a pairing term in order to get the appropriate symmetry breaking. Putting the chemical potential $\epsilon_0$ to zero, the simplest possible action for an extended kink is,
\be{xilag} 
{\cal L}_{\xi \theta b} =  \frac 1 {2\pi} \epsilon^{\mu\nu} \partial_\nu\theta \, \left[ b_\mu  + \bxi  i\partial_\mu  \xi\right]  -{\cal H}_\xi   - b_\mu j^\mu_k
\end{align}
with
\be{xiham}
{\cal H}_\xi = \frac{1}{2\pi} \bxi \bigl(M^2 - \partial^2_x\bigr)  \xi +
\frac{ \delta M} {4\pi} \left[ e^{i\theta} \xi i \partial_x \xi + e^{-i\theta} \bxi i\partial_x \bxi \right],
\end{align}
where the mass parameter $M$ and the pairing strength $\delta$, are phenomenological parameters. 

We can simplify this Hamiltonian by performing a rotation of the fields: 
\begin{align}
\label{rotations}
\xi &\rightarrow e^{-i\theta/2}\xi &
\xi^\dag &\rightarrow e^{i\theta/2}\xi^\dag \ .
\end{align}
This will transform the Hamiltonian \eqref{xiham} to $\cal{H}_\xi = (\xi^\dag,\xi) \bar{\cal{H}} (\xi,\xi^\dag)^\text{T}$ with
 \begin{align}
 \label{xihammatrot}
 {\cal \bar{H}}_\xi =
 \left(\begin{array}{cc} M^2-(\partial_x-\frac{i}{2} \partial_x\theta)^2 &  \delta M i\partial_x \\ 
  \delta M i\partial_x & -M^2+(\partial_x+ \frac{i}{2} \partial_x\theta)^2 
   \end{array} \right).
\end{align}

Next, we expand the quantum field as
$\xi(x,t) = \sum_n (e^{-iEt} u^*_n(x) c^\dag_n + e^{iEt} v^*_n(x) c_n)$,
which yields the following BdG equations for the eigenfunctions $u(x)$ and $v(x)$,
\begin{align}
\label{xieigen}
\Bigl( (\partial_x+\frac{i}{2}\partial_x\theta)^2 + E \partial_x \theta - M^2 \Bigr) u(x)
+\delta M i\partial_x v^*(x) = 0 \\ 
\Bigl( (\partial_x-\frac{i}{2}\partial_x\theta)^2 - E \partial_x \theta - M^2 \Bigr) v^*(x)
- \delta M i\partial_x u(x) = 0
\nonumber
\end{align}
In the limit $\delta\rightarrow 0$ and under the assumption that
$\theta$ varies slowly (i.e, we assume $\partial^2_x\theta$ and $(\partial_x\theta)^2$ to be small)
we obtain the following equation for $u(x)$ 
\begin{equation}
\label{finalu}
\left[\partial_x^2 + (\partial_x\theta)i\partial_x + E(\partial_x\theta) - M^2\right]u(x)=0,
\end{equation} 
which is \eqref{secondor} in the limit where the energy $E$ is small compared to $M$.
As expected there is no continuous component in the spectrum, and the low energy part of the
spectrum compares well with the full model with suitable adjustment of the model parameters. In particular, we should set $M^2=\xi_0^{-2}=4\Delta^2/v_F^2$ , $E=\tilde{\epsilon}=2\epsilon/v_F$. The requirement that $E \ll M$ then translates to $\epsilon \ll \Delta$, that is, for energies well below the gap, which is consistent with a zero energy bound state.

To actually {\em derive} the effective Lagrangian \pref{xithblag} one should  integrate out the 
 high energy modes. This would not only give expressions for the effective parameters, but also provide an ultraviolet cutoff that  would define the region of validity of the effective model. We have not been able to do this in a controlled way, but in  Appendix \ref{efftheoryapp} it is shown, by manipulating path integral expressions, how the crucial kinetic term
$ \half \epsilon^{\mu\nu} \partial_\nu\theta \bxi  i\partial_\mu\xi$  can arise from the microscopic description. 

Finally we note that the extension of the topological theory \eqref{xithblag} to the model Lagrangian \eqref{xilag} for the 
sub-gap regime, is reminiscent of the extension, proposed in \oncitep{hansson14up} of the 2D topological theory in 
\oncitep{hansson12} In both cases the models are constructed using phenomenological and heuristic arguments, and it remains 
a theoretical challenge to  find general methods to describe localized fermionic zero modes in the general context of
topological field theory. 

\section{Concluding remarks}
\label{sec:conclusion}
In this paper we studied several models for trivial and topological superconducting wires in one dimension. More specifically, we investigated the properties of $\pi$-junctions, and in particular those where the phase of the order parameter winds an angle $\pi$ over the junction, corresponding to a system in symmetry class D.
For this more general case, we find that there is no topologically protected zero energy mode associated with a $\pi$-junction. Rather, local breaking of the PTRS by means of the complex winding of the order parameter can shift the energy of the bound state in the junction region away from zero energy. This symmetry breaking is not allowed in class BDI, where, as a consequence, the bound state is topologically pinned to zero energy.
We demonstrated that the low energy bound states in some specific cases can be obtained analytically and showed that these results agree well with numerical calculations. 
Most importantly, we discussed how our results might be used to obtain a bulk probe - in contrast to the common method of probing the edges - to  distinguish a topological wire from a trivial one, and suggested some experimental approaches to this end. 
Finally we constructed a low energy field theory with a topological term describing itinerant $\pi$-junctions, and discussed its relation to theories in two dimensions.

\vskip 3mm
{\em Acknowledgements}. 
C.S. thanks S. Abay Gebrehiwot, M. Leijnse, H.Q. Xu and C. Yu for interesting discussions and hospitality. T.H.H. thanks F. von Oppen for a useful discussion.
This research was sponsored, in part, by the Swedish research council. J.C.B. acknowledges funding from the ERC synergy grant UQUAM.

\appendix

\section{Topology of the $p$-wave superconductor modes.}
\label{apptoplin}

Here, we discuss the topological properties of the various 
models we consider in this paper. To set the scene, we start  by
recalling the topological properties of the Kitaev chain, see Ref. [1].

\begin{figure}[h]
\includegraphics[width=\columnwidth]{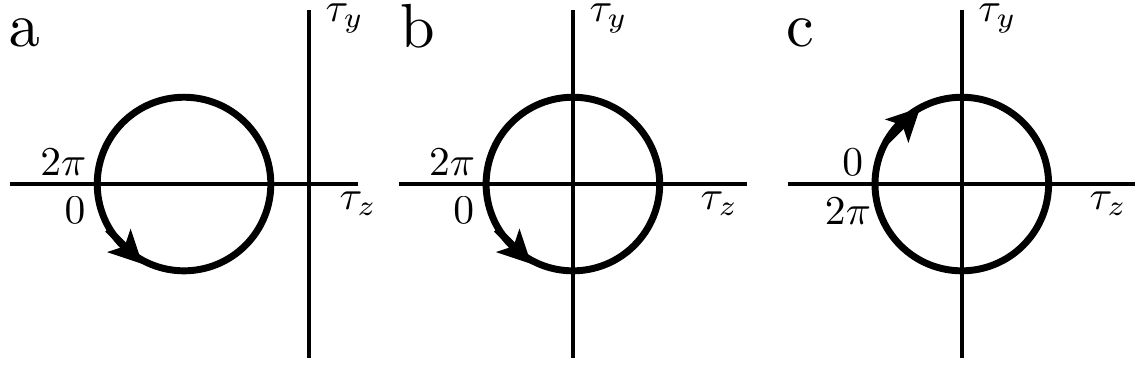}
\caption{Winding numbers $\nu$ of $\vec{d}(k)$ for the full Kitaev chain, in
(a) trivial phase with $\nu=0$, for $0< t < \mu/2$, $\Delta > 0$,
(b) topological phase with $\nu=1$ for $\mu=0$, $0 < t = \Delta$ and
(c) topological phase with $\nu=-1$ for $\mu=0$, $0 < t = -\Delta$.
The arrows denote the direction in which $k$ increases.}
\label{fig:winding-full}
\end{figure}

Consider the model \eqref{kitaevstart} and  assume that $t$ and $\Delta$ 
are both real, so that the Hamiltonian belongs to symmetry class BDI.
The topological invariant takes the form of a winding number\cite{Schnyder2008}, and 
to show this in the present case, we write the $k$-space Hamiltonian~\eqref{kitaevbdg} as
\begin{equation}
\mathcal{H}_K(k) = \vec{d} (k) \cdot \vec{\tau} \, , \label{eq:dvec}
\end{equation}
with $\vec{\tau} = (\tau_x, \tau_y, \tau_z)$.
For models in class BDI, one can choose a basis such that one of the components of the
vector $\vec{d}$ is zero, say $d_x = 0$. The energy is given by
$\epsilon (k) = \pm | \vec{d}(k) |$, which means that for a gapped system, we have
$\vec{d}^2 (k) > 0$. Hence, the winding number $\nu$ around the origin of the curve in
$(\tau_y,\tau_z)$-space
(\emph{i.e.}, the space of Hamiltonians) swept out by
$\vec{d} (k)$ as $k$ sweeps through the full Brillouin zone is well defined. This
winding number is the topological invariant characterizing the different phases.
For the Kitaev chain   we have
$\vec{d} (k) = (0,-\Delta \sin (k),-\mu/2-t \cos(k) )$, and in 
Fig.~\ref{fig:winding-full}, we (schematically) show the curve $\vec{d}(k)$ in the trivial phase,
with winding $\nu=0$, and the two different topological phases, with winding
$\nu = \pm 1$.

Next we turn to the linear model ${\cal H}_{\rm Lin}$ in  Eq.~\eqref{linsc}.  Assuming that $\Delta$ is real and constant,
 the momentum space version of the Hamiltonian \eqref{Hcurl}  is again of the form \eqref{eq:dvec}, with 
 $\vec{d}(k) = (0,-2\Delta, v_F k)$.

\begin{figure}[h]
\includegraphics[height=3cm]{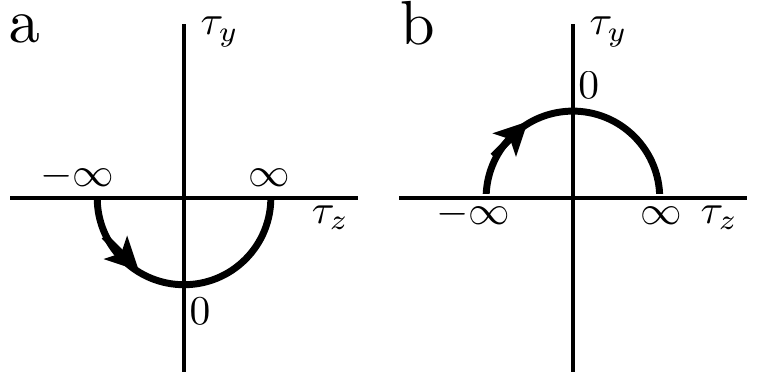}
\caption{`Winding' of the vector $\hat{d}(k)$ for the linearized model \eqref{Hcurl}, for
(a) $\Delta >0$ and (b) $\Delta < 0$.}
\label{fig:winding-lin}
\end{figure}

Since  the $k$-space is not compact, it is possible that
the curve swept out by $\hat{d}(k)=\vec{d} (k)/|\vec{d} (k)|$ (the normalization is needed to 
obtain finite limits and is valid as long as the Hamiltonian is gapped) as $k$ goes from $-\infty$ to $\infty$ is not closed. 
This is indeed what we find in Fig.~\ref{fig:winding-lin}, where we depict the two cases $\Delta = \pm 1$.
Despite that we can not define a winding
number for neither of the values $\pm \Delta$,  we can still 
consider the {\em difference in winding number } $\delta \nu$
between the two cases, which gives $|\delta \nu| = 1$. Therefore, we  expect a
zero energy bound state at a boundary between two regions with
$\Delta = \pm 1$ respectively, even in the linearized model. We stress, that although
that this argument in not rigorous, it is nevertheless true, and in the main text 
we showed that  the
analytic form of the zero mode of the linearized model of
Ref.~\onlinecite{Takayama:1980zz} accurately describes the DZM in the
junction of the full Kitaev chain.

\begin{figure}[h]
\includegraphics[height=2cm]{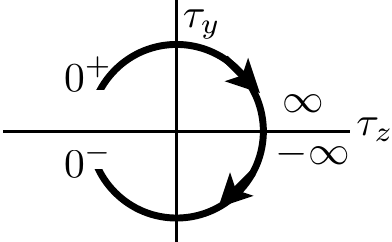}
\caption{The `winding' of the $\hat{d}(k)$-vector corresponding to the linearized model
${\mathcal H}_v$ before the regularization which
removes the discontinuity at $k=0$. The arrows indicate the direction in which
$k$ increases. We have used $\Delta>0$. As discussed in the text, the gap between the points
$0^+$ and $0^-$ is closed if the dispersion relation is smoothened at $k=0$.}
\label{fig:winding-lin2}
\end{figure}

We now turn to the alternative  linearized model given by
${\mathcal H}_v$ in  Eq.~\eqref{vmodel}. 
Here the $k$-space is  again not compact, and there is also a discontinuity at $k=0$.
The first issue is remedied by identifying the
points at $\pm\infty$ (which amounts to considering the $a\rightarrow 0$ limit of the lattice model). 
To deal with the second, we note that for this model,
$\vec{d}(k)  = \frac{1}{2}( 0 , \Delta {\rm sgn} (k) , -\bar{\mu} + v_F |k|)$,
and in  Fig.~\ref{fig:winding-lin2}
we show the corresponding `winding' of the vector $\hat{d}(k) = \vec{d}/|\vec{d}|$,
in the case $\Delta > 0$.
Even when identifying the points at $k = \pm\infty$, the curve is not continuous, but with a 
regularization that smoothens out the singularity in the V-shaped band, by replacing 
the factor ${\rm sgn}$ by a continuous odd function that  rapidly changes sign  around $k=0$, 
the $d$-vector will  will be continuous, and the winding number will be well defined.
This concludes the demonstration  of the existence of 
a linearized continuum model with topological properties identical to that of the 
Kitaev chain.

\section{Topological aspects of the $s$-wave paired models}
\label{sec:AppTop}
In this appendix, we discuss the topological properties of the full and linearized $s$-wave models. Due to the extra spin degree of freedom in these models, the winding arguments used for the $p$-wave superconductors are not directly applicable and another method of topological classification must be used. We will use the method outlined in Ref.~\onlinecite{SauBDI}.

We begin with the full $s$-wave model, given by \eqref{sHam}. Assuming a real and constant order parameter, the corresponding $k$-space Hamiltonian can (in suitable units) be written as 

\begin{align}
\label{HkS}
H^S(k) = (k^2-\bar{\mu}) \tau_z s_0 -  \Delta \tau_y s_y
\end{align}
where the Pauli matrices $\tau_i$ and $s_i$ act in particle-hole space and spin-space respectively. This Hamiltonian belongs to symmetry class BDI, meaning PTRS $\mathcal{T}^2 = +1$ and PHS $\mathcal{C}^2 = +1$. These operators are in our chosen basis given by $\mathcal{T} = \tau_0 s_0K$ and $\mathcal{C} = \tau_x s_0 K$, with $K$ denoting the complex conjugation operator.

To investigate the topological properties of this Hamiltonian, we write it in the form

\begin{align}
\label{Hkmat}
H^S(k) = 
 \begin{pmatrix}
  H_0(k) & \hat{\Delta} \\
  \hat{\Delta}^\text{T} & -H_0(k)
 \end{pmatrix},
\end{align}
where the matrix structure is in particle-hole space, $H_0(k) = (k^2-\bar{\mu})s_0$ and $\hat{\Delta} = \Delta i s_y$. We note that the latter term is real and has the property $\hat{\Delta}^\text{T} =-\hat{\Delta}$. By a unitary transformation with $U = \exp(-i(\pi/4) \tau_y s_0)$, the matrix in equation \eqref{Hkmat} can be rotated into 

\begin{align}
\label{Hkmatdiag}
UH^S(k)U^\dag = 
 \begin{pmatrix}
  0 & A(k) \\
  A(-k)^\text{T} & 0
 \end{pmatrix},
\end{align}
with $A(k) = H_0(k) + \hat{\Delta}$. 

Next, we note that $\text{Det}(H^S(k)) = \text{Det}(UH^S(k)U^\dag) = \text{Det}(A(k))\text{Det}(A(-k)^\text{T})$ so that if $H^S(k)$ is gapped for all $k$, \ie $\text{Det}(H^S(k)) \neq 0$, the determinant of $A(k)$ can not vanish either. This allows us to define $z(k) =\exp(i\theta(k))= \text{Det}(A(k))/|\text{Det}(A(k))|$ for gapped Hamiltonians $H^S(k)$. One may then show that 

\begin{align}
\label{zk}
z(k) = \text{sgn}(\Delta^2 + (k^2-\bar{\mu})^2).
\end{align}

For the $s$-wave Hamiltonian \eqref{HkS}, which is real and gapped for all $k$, $\bar{\mu}$ and finite $\Delta$, the determinant is real and non-vanishing. Then $z(k)$ is well defined and is equal to $+1$, independently of any compactification of $k$-space (which is needed for any well defined topological invariant), rendering the model topologically trivial.

We next turn our attention to the linear $s$-wave model. To derive it, we apply the linearization scheme described in section \ref{sec:linear} to \eqref{sHam} and again assume a real order parameter which gives us the following Hamiltonian:

\begin{align}
\label{linscS}
& H^S_{\rm Lin} =  \int dx \sum_\sigma((-iv_F \varphi_{\sigma,+}^\dag\partial_x \varphi_{\sigma,+}  + iv_F \varphi_{\sigma,-}^\dag\partial_x \varphi_{\sigma,-}) + \notag \\
&  \Delta(x)(\varphi_{\uparrow,+} \varphi_{\downarrow,-} + \varphi_{\uparrow,-} \varphi_{\downarrow,+}+\varphi^\dag_{\downarrow,+} \varphi^\dag_{\uparrow,-} + \varphi^\dag_{\downarrow,-} \varphi^\dag_{\uparrow,+}).
\end{align}

We write this as

\begin{align}
\label{HSLin}
H^S_{\rm Lin} = \int dx \Psi^\dag {\cal H^S}_{\rm Lin}(x)\Psi
\end{align}

with 
\begin{align}
\label{HSLinMat}
{\cal H}^S_{\rm Lin} = \begin{pmatrix}
  -iv_F\partial_x & 0 & 0 & -\Delta \\
  0 &-iv_F\partial_x & \Delta & 0 \\
  0  & \Delta  & iv_F\partial_x & 0  \\
  -\Delta & 0 & 0 & iv_F\partial_x
 \end{pmatrix},
\end{align}

and the basis

\begin{align}
\Psi = (\varphi_{\uparrow,+},\varphi_{\downarrow,+},\varphi^\dag_{\uparrow,-},\varphi^\dag_{\downarrow,-})^{\rm T}.
\end{align}

The matrix in equation \eqref{HSLinMat} looks very much like two separate blocks of the linear $p$-wave superconductor which seems a bit troublesome since we know that the linear $p$-wave model host zero modes. That would imply that the linear $s$-wave model {\it also} would host zero modes, which would contradict our findings in this paper. 

One may suspect that the appearance of two $p$-wave models is incidental, and that by adding corrections to the linearization this illusion is shattered. This suspicion is indeed justified since, as  we now show, the linear $s$-wave superconductor in fact is topologically trivial. 

The corresponding $k$-space Hamiltonian is given by
\begin{align}
\label{HkSLin}
H^S_{\rm Lin}(k) = v_F k \tau_z  s_0 + \Delta \tau_y s_y
\end{align}

where the Pauli matrices $\tau_i$ and $s_i$ now act in right-left space and spin-space respectively. This Hamiltonian also belongs to class BDI. In our basis the particle-hole (now right-left) symmetry operator is given by $\mathcal{C} = \tau_z  s_0 K$ and the pseudo time reversal symmetry operator is $\mathcal{T} = \tau_x s_x K$. We rotate the Hamiltonian with the unitary matrix $U = \exp(-i(\pi/4) \tau_x s_x)$, giving us a structure like \eqref{Hkmatdiag} but now with $A(k) =iv_F k s_x +i\Delta s_y$. One may then, as above, define $z(k)$ which in this case turns out to be $z(k)=\text{sgn}(v_F^2k^2+\Delta^2)=+1$ for all $v_F$, $k$ and finite $\Delta$. As was the case in Appendix~\ref{apptoplin},
k-space is not compact. Regardless of this issue, $z(k)$ can never wind.

Thus we can conclude that both the full and linear $s$-wave superconductors are trivial,
and hence that the zero modes these models exhibit are not topologically protected.

\section{Origin of the term $  \epsilon^{\mu\nu} \partial_\nu\theta \bxi  i\partial_\mu\xi$  } \label{efftheoryapp}
Starting from the original Lagrangian \pref{diraclag}, we present an argument for how the kinetic term  $ \frac 1 {2\pi} \epsilon^{\mu\nu} \partial_\nu\theta \bxi  i\partial_\mu\xi$
can appear in an effective Lagrangian. Although, as already emphasized in the main text, several of the steps in the below derivations are based on unproven assumptions, the emergence of the kinetic term is far from obvious, and this indicates that a more rigorous proof along these lines might be possible. 

The starting point is the partition function,
\be{partf}
Z[\theta,g] = \int {\mathcal D}[\psib,\psi] e^{i\int d^2x\, {\cal L} (\psib,\psi, \theta ) }.
\end{align}
The strategy is  to change fermionic variables in such a way that the high energy part of the spectrum can still be bosonized and integrated out, as in the previous section, while the the low lying fermion spectrum will be captured by a Lagrangian like \pref{xithblag}. To this end, we shall use the following identity,
\be{pathid}
&\int {\mathcal D} [a_\mu ] {\mathcal D}[\bxi,\xi] e^ { i\int d^2x\,  \left[  a^\mu (\bxi p_\mu \xi - \psib \gamma_\mu \psi )
 - {\cal H}_\xi \right] } \nonumber \\
&= \int {\mathcal D} [a_\mu  ] e^ { i\int d^2x\,   a^\mu j_\mu   + \half \mathrm{Tr}\ln ( H + a^\mu p_\mu)  } \nonumber \\
&=e^{i {\mathcal F}[j_\mu] },
\end{align}
where $p_x =-i \partial_x$, $j_\mu = \psib \gamma_\mu \psi$ and  ${\cal H}_\xi$ is an Hamiltonian that we shall assume to be quadratic in the fields and $H$ is
the corresponding operator acting on the Nambu spinors.

To derive the last line in \pref{pathid} we first calculate the lowest order by expanding the logarithm and evaluating the trace (which is over both space and Nambu indices). The resulting integrals are not convergent in the ultraviolet since there is no time derivative, so we must introduce a cutoff energy scale $\Lambda$. The resulting effective 
theory is only to be applied below this scale.
Note that  there is no gauge invariance related to the auxiliary field $a$ since it does not couple to a conserved current. Taking for ${\cal H}_\xi$ the 
expression \pref{xiham} a straight forward calculation gives  $\mathrm{Tr}\ln ( H + a^\mu p_\mu)  = c_0  a_0^2 +  c_1  a_1^2 + \dots$ where we  omitted all 
higher derivative terms. The explicit expressions for the coefficients in terms of $\delta$, $M^2$ and $\Lambda$ are not particularly illuminating. 
Substituting this in the second line of \pref{pathid} and integrating over $a$, we retain the third line with 
\be{auxfun}
 {\mathcal F}[j_\mu]  = \tilde c_0 \, j_0 ^2 +   \tilde c_1 \, j_1 ^2 + \dots \,.
 \end{align}
Before inserting the identity \pref{pathid} in the path integral \pref{partf}, we perform the chiral rotation,
\be{chirrot}
\psi \rightarrow e^{\frac i 2 \vartheta(x) \gfem } \psi 
\end{align}
under which the Lagrangian \pref{diraclag} becomes,
\be{rotmodl}
{\cal L} = \psib \left( i \slashed \partial - \pi \slashed j_k - g(x) \right) \psi \, .
\end{align}
Putting this together, we get the following representation for the partition function,
\be{partf2}
Z[\theta,g] = \int  {\mathcal D} [a_\mu ] {\mathcal D}[\bxi,\xi]  {\mathcal D}[\psib,\psi] e^{i S[a, \bxi,\xi, \psib,\psi ; \theta]} \ ,
\end{align}
\be{extact}
S &= \int d^2x\, [ \psib \left( i \slashed \partial  - \pi \slashed j_k  - \slashed a - m \right) \psi   \\ 
&+ \tilde c_0 (\psi\gamma_0\psi)^2  + \tilde c_1 (\psi\gamma_1\psi)^2     
- a^\mu \bxi i\partial_\mu \xi  - {\cal H}_\xi ] \ , \nonumber
\end{align}
where we put $g(x) = m$ to connect to the previous discussion about the kink solutions. Next we make a shift $a^\mu \rightarrow a^\mu -\pi  j^\mu_k$, to rewrite the action as 
\be{extact2}
S = \int d^2x\, [ {\cal L}_f  - a^\mu \bxi i\partial_\mu \xi   
+ \pi j_k^\mu\,  \bxi i\partial_\mu\xi   - {\cal H}_\xi ]  \ ,
\end{align}
where
\be{psilag}
{\cal L}_f =  \psib \left( i \slashed \partial  - \slashed a - m \right) \psi +\tilde c_0 (\psi\gamma_0\psi)^2  + \tilde c_1 (\psi\gamma_1\psi)^2     
\end{align}
is very similar to the massive Thirring model. The $\psi$-field can now be integrated to give an effective Lagrangian, for the $a_\mu$ field. Using the gauge invariance of \pref{psilag} we get 
\be{effalag}
{\cal L}_{eff} (a) = - \frac 1 {\tilde m^2}F^2 \dots
\end{align}
where $F_{\mu\nu}$ is the field strength for the potential $a_\mu$ and  $\tilde m$ a dimensional constant that depends both on $m$ and, via the coefficients $\tilde c_0$ and $\tilde c_1$, on $\delta$, $M^2$ and $\Lambda$. 
Finally, we can integrate the  vector field $a_\mu$ to get the desired effective action for the $\xi$-field, 
\be{xithblag2}
{\cal L}_{\xi \theta b} = \half \epsilon^{\mu\nu} \partial_\nu\theta \, (b_\mu  + \bxi i\partial_\mu \xi)     - {\cal H}_\xi - b_\mu j^\mu_k + \dots
\end{align}
where we also used the constraint \pref{kinkcurr} to express $j_k^\mu$ in terms of $\theta$, and where the dots indicate both neglected higher derivative terms in the quadratic action, and interaction terms resulting from integrating the $a_\mu$ field. All the steps glossed over above can be performed, at least to low order in perturbation theory. The main question is  however not technical, but rather what principle should be used to determine $  {\cal H}_\xi $. A possible approach is to choose the parameters in $  {\cal H}_\xi $ so to minimize the size of the leading corrections due to higher derivative terms and induced interactions.


\end{document}